\documentclass[%
 reprint,
 amsmath,amssymb,
 aps,
prb,
floatfix,
]{revtex4-2}

\usepackage{graphicx}% Include figure files
\usepackage{dcolumn}% Align table columns on decimal point
\usepackage{bm}% bold math
\usepackage{mathrsfs}
\begin{document}

\preprint{APS/123-QED}

\title{The 6H-Perovskite Dimer Lattice with Antiferromagnetic
Interactions: Ba$_3$ARu$_2$O$_9$}

\author{Daniel M. Pajerowski}
\email{pajerowskidm@ornl.gov}
\affiliation{Neutron Scattering Division, Oak Ridge National Laboratory, Oak Ridge, Tennessee, USA}
\author{David A. Dahlbom}
\affiliation{Neutron Scattering Division, Oak Ridge National Laboratory, Oak Ridge, Tennessee, USA}
\author{Daniel Phelan}
\affiliation{Materials Science Division, Argonne National Laboratory, Lemont, IL 60439, United States of America}
\author{Yu Li}
\affiliation{Materials Science Division, Argonne National Laboratory, Lemont, IL 60439, United States of America}
\author{Alexander I. Kolesnikov}
\affiliation{Neutron Scattering Division, Oak Ridge National Laboratory, Oak Ridge, Tennessee, USA}

\date{\today}% It is always \today, today,

\begin{abstract}
We investigate the magnetic behavior of the 6H-perovskite dimer lattice Ba$_3$Zn$_{1-x}$Ca$_x$Ru$_2$O$_9$ using analytical theory, density functional theory, inelastic neutron scattering, and modeling of historical magnetization and neutron-scattering data. A dimer mean-field theory built upon classical Luttinger--Tisza analysis generates a phase diagram revealing a transition from a nonmagnetic singlet to a finite-moment ground state as interdimer couplings increase. A (generalized) linear spin-wave theory captures multiplet mixing, excitation gap closing, and fluctuation-induced moment suppression. Density functional theory on select compounds and neutron spectroscopy on dilute Ba$_3$Zn(Ru$_{1-x}$Sb$_x$)$_2$O$_9$ confirm the exchange hierarchy, enabling quantification of previously published experiments within this framework. Our results identify three mechanisms for magnetic moment suppression—quantum fluctuations, ligand hybridization, and nonmagnetic-singlet/magnetic-multiplet mixing.
\end{abstract}

\maketitle

%\tableofcontents

\section{\label{sec:intro}Introduction}

\begin{figure}[thbp]
  \centering
  \includegraphics[width=\columnwidth]{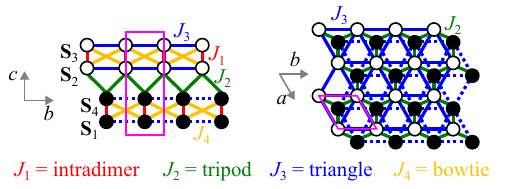}
  \caption{(Color online) Geometry and exchange interactions of the 6H-perovskite, AB-stacked dimer
triangular lattice. The unit cell is outlined in magenta and contains two dimers, one in each layer. The left panel
shows a side view of the structure, with white and black circles
representing magnetic dimers in alternating triangular layers along the
hexagonal \emph{c}-axis; nonmagnetic ions are omitted for clarity. The
right panel shows the top-down view, where the dimer units form an
AB-stacked triangular lattice. Four distinct antiferromagnetic exchange
interactions are indicated: intradimer ($J_1$, red), tripod ($J_2$, green),
triangle ($J_3$, blue), and bowtie ($J_4$, yellow). The black circles are
connected with dashed lines for $J_3$.}
  \label{fig:lattice}
\end{figure}

Hexagonal perovskites can host regular lattices of magnetic ions and
exhibit a variety of geometries within that motif, giving rise to
diverse magnetic ground states. A recent review article shows the rich
chemistry and physics of triple perovskites~\cite{De2024}. Here, we focus on
the triple perovskite with 6H-type stacking (in Ramsdell
notation)~\cite{Burbank1948}, in which there are systems with two vertically offset 
magnetic dimers per unit cell that are each spaced by nonmagnetic
monomers. The resulting triangular layers of dimers are arranged in an
AB stacking pattern, with each dimer positioned above and below the
center of a triangle in the vertically adjacent layer. This lattice
structure, along with the four shortest exchange pathways, is
illustrated in Figure~\ref{fig:lattice} and the Hamiltonian is given by:
\begin{eqnarray}
H &=& J_{1} \sum_{\langle i,j \rangle_1} \mathbf{S}_i \cdot \mathbf{S}_j
   + J_{2} \sum_{\langle i,j \rangle_2} \mathbf{S}_i \cdot \mathbf{S}_j \nonumber \\
  && + J_{3} \sum_{\langle i,j \rangle_3} \mathbf{S}_i \cdot \mathbf{S}_j
   + J_{4} \sum_{\langle i,j \rangle_4} \mathbf{S}_i \cdot \mathbf{S}_j
\label{eq:Hamiltonian}
\end{eqnarray}
where the indices 1--4 on exchanges refer to the bond types shown in
Figure~\ref{fig:lattice}.

Among materials that realize this lattice geometry, we focus on
Ba$_3$Zn$_{1-x}$Ca$_x$Ru$_2$O$_9$, which has been the subject of experimental studies, revealing subtle structural changes with composition $x$ that arise from the different ionic radii of Ca$^{2+}$ and Zn$^{2+}$ (Ca$^{2+}$ is approximately $30\%$ larger)~\cite{Shannon1976} and that significantly influence its magnetic properties. Because Ca$^{2+}$ and Zn$^{2+}$ are isovalent, these structural changes primarily affect the magnetic interactions between Ru$^{5+}$, $S=\tfrac{3}{2}$ ions. As $x$ decreases, the interdimer separation decreases, and the system transitions from a regime consistent with a gapped singlet ground state of weakly coupled antiferromagnetic dimers to one whose magnetic behavior has not yet been quantitatively modeled, but is inconsistent with isolated dimers. The ground state of Ba$_3$ZnRu$_2$O$_9$ (BZRO) is anomalous in that it exhibits a finite (gapless?) magnetic response down to $37~\mathrm{mK}$, without typical experimentally observed signatures of long-range dipolar order such as magnetic neutron diffraction, a peak in the specific heat, or a cusp in the magnetic susceptibility~\cite{Terasaki2017}. The triangular plane motif, the large Curie-Weiss temperature with no long-range order detected, and unquantified low-temperature magnetic response were therefore suggested to indicate spin-liquid behavior in BZRO.\cite{Terasaki2017}

A rich literature exists on
Ba$_3$ARu$_2$O$_9$ compounds,
encompassing both magnetic and non-magnetic A-site ions. The earliest
studies treated these systems with diamagnetic A-sites as consisting of
isolated antiferromagnetically coupled $S = \tfrac{3}{2}$
Ru$^{5+}$--Ru$^{5+}$ dimers. For example,
magnetic susceptibility data for A = Ca, Cd, Sr, and Mg were modeled
using non-interacting dimers, yielding semi-quantitative agreement with
fitted $J_1$ values of 29.3, 29.8, 29.8, and 23.8 meV,
respectively~\cite{Darriet1976}. To improve agreement for A = Ca, a modified
Hamiltonian including biquadratic exchange was introduced to adjust the
multiplet level spacings~\cite{Drillon1977}. Inelastic neutron scattering (INS) of
Ba$_3$CaRu$_2$O$_9$ (BCRO) was
modeled using a non-interacting dimer model to extract
$J_1$=26 meV from a magnetic peak position at one momentum
point, although the non-resolution-limited linewidth and changing peak
position with momentum was not quantified~\cite{Darriet1983}.

Neutron diffraction measurements have been reported for A = Ni, Co, and
Zn, with magnetic Bragg peaks observed for A = Ni and Co (magnetic
A-sites), but no magnetic Bragg peaks for A = Zn~\cite{Lightfoot1990}. The absence of
magnetic scattering in the Zn compound was attributed to insufficiently
low temperature ($T = 5~\mathrm{K}$), referencing earlier Mössbauer spectroscopy
that hinted at ordering just below 4.2 K~\cite{Fernandez1980}. Those Mössbauer
measurements were conducted at $T = 4.2$ K for A = Ca, Sr, Cd, Mg, Zn, Co,
Ni, and Cu showed no line-splitting for Ca, Sr, and Cd, consistent with
a singlet ground state. In contrast, magnetic splitting was observed for
Co, Ni, and Cu (magnetic A-sites) consistent with known magnetic
ordering, as well as for Zn and Mg (diamagnetic A-sites)---raising
questions about the magnetic ground states of the latter compounds and
suggesting either static order or slow spin relaxation on the Mössbauer
timescale.

More recently, neutron powder diffraction on
Ba$_3$ZnRu$_{2-x}$Ir$_{2x}$O$_9$ ($x=0,1,2$)
and BCRO revealed no magnetic Bragg peaks down to 5 K~\cite{Beran2015} and 1.6
K~\cite{Senn2013}, respectively. A separate study comparing dilute A-site
substitution (2\% Co, Ni, or Cu) in BZRO and BCRO concluded that the
dilute ions in the substituted Zn compound might represent a spin-liquid
analogue of the Kondo effect~\cite{Yamamoto2018}. Investigations into Nb
substitution on the Ru site in
Ba$_3$Ca(Ru$_{1-x}$Nb$_x$)$_2$O$_9$
and
Ba$_3$Sr(Ru$_{1-x}$Nb$_x$)$_2$O$_9$
were modeled with $S = 1$ (not the Hund $S = \tfrac{3}{2}$) for the
NbRuO$_9$ dimers in the Ca system~\cite{Nishihara2021}, though later
work on that compound proposed a more conventional $S = \tfrac{3}{2}$ model with
internal magnetic fields~\cite{Ochiai2024}. Finally, a resonant inelastic x-ray
scattering (RIXS) study of BZRO tested the possibility of an
orbital-selective S = 1 state, but concluded that the system hosts $S =
\tfrac{3}{2}$ antiferromagnetic dimers~\cite{Hayashida2025}.

To better understand the anomalous magnetic behavior of
Ba$_3$Zn$_{1-x}$Ca$_x$Ru$_2$O$_9$,
and to explore the broader physics of this dimer lattice, we investigate
the ground-state phase diagram as a function of antiferromagnetic
exchange interactions.
While motivated by the unexplained observations,
our goal is also to characterize the possible phases that can arise in
this geometry more generally, including regimes not yet realized in
experiment. Our approach combines classical Luttinger--Tisza (LT)
analysis, dimer mean-field theory incorporating quantum degrees of
freedom, and linear spin-wave theory (LSWT) to evaluate the effects of
quantum fluctuations. These methods are used to construct a global phase
diagram in Section~\ref{sec:phasediag}, which is then compared to experimental observations in Section~\ref{sec:BZCRO}. By
digitizing previously published data, performing density functional
theory (DFT) calculations, and analyzing both previously reported and
newly collected neutron spectroscopy results, we assign different
compositions of
Ba$_3$Zn$_{1-x}$Ca$_x$Ru$_2$O$_9$
to specific regions of the phase diagram and examine how the underlying
magnetic correlations evolve with substitution. 

Our analysis suggests that
the moment suppression observed in this family of compounds, as well as the complete 
lack of detectable magnetic order in the case of BZRO, can be attributed to three
mechanisms. When interdimer exchanges are small, moments are
substantially suppressed due to localized quantum fluctuations on the $J_1$ bonds: the ground state is approximately a product of nonmagnetic singlets or, for slightly larger interdimer exchange, singlets mixed with magnetic multiplets. When interdimer exchanges are large, leading to the emergence of finite magnetic dipoles on each site, zero-point corrections calculated using standard LSWT suggest a significant moment reduction due to the collective quantum fluctuations typically observed in (frustrated) antiferromagnetic systems. On top of these mechanisms, ligand hybridization acts to reduce the local moments throughout the phase diagram. 

\section{\label{sec:phasediag}Theoretical phase diagram}

\subsection{\label{sec:dimer}Dimer Mean-field Theory}

To generate a phase diagram for the 6H-perovskite dimer lattice with antiferromagnetic
interactions---including the possibility of a dimer forming a singlet
ground state when energetically favorable---we implement a dimer
mean-field theory~\cite{White2007} for the ground state from the LT analysis in Appendix~\ref{sec:LT_appendix}. In this approach, each magnetic dimer is locally
treated as a quantum mechanical two-site system, while the couplings
between dimers are approximated at the mean-field level. The resulting
Hamiltonian includes an exact intradimer exchange term and mean-field
approximations for all interdimer couplings. Explicitly, within the
context of the previous definitions, the mean-field Hamiltonian for one symmetrically distinct dimer takes
the form:

\begin{eqnarray}
H &=& \mathbf{J}_{1}(\mathbf{k})\, \mathbf{S}_{2} \cdot \mathbf{S}_{3}
  + \mathbf{J}_{2}(\mathbf{k}) \left[ \mathbf{S}_{2} \cdot \langle \mathbf{S}_{3} \rangle \right] \nonumber \\
  && + \mathbf{J}'_{2}(\mathbf{k}) \left[ \langle \mathbf{S}_{2} \rangle \cdot \mathbf{S}_{3} \right]
  + \mathbf{J}_{3}(\mathbf{k}) \left[ \mathbf{S}_{2} \cdot \langle \mathbf{S}_{2} \rangle \right] \nonumber \\
  && + \mathbf{J}_{3}(\mathbf{k}) \left[ \mathbf{S}_{3} \cdot \langle \mathbf{S}_{3} \rangle \right]
\label{eq:Hk}
\end{eqnarray}
where $\mathbf{S}_{2}$ and $\mathbf{S}_{3}$ label the two spins within the
dimer, and the angle brackets denote expectation values taken over
neighboring dimers. The momentum dependence reflects the Fourier
structure of the exchange couplings. Note that the lowest energy mode obtained from the LT analysis involves only $\mathbf{S}_2$ and $\mathbf{S}_3$. Numerical details are in Appendix~\ref{sec:dimer_appendix}.

\begin{figure}[thb]
  \centering
  \includegraphics[width=\columnwidth]{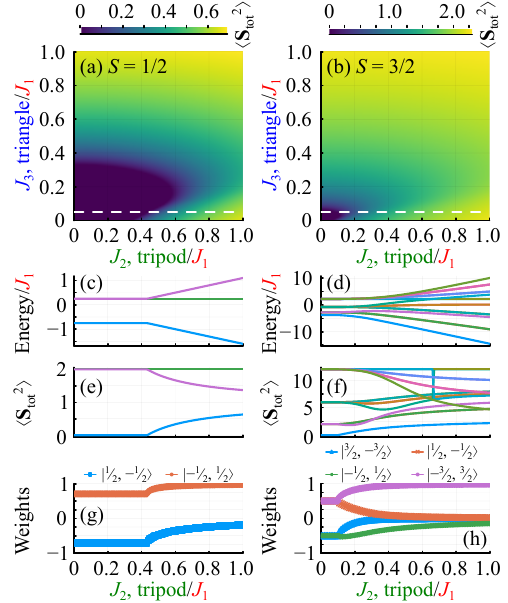}
  \caption{(Color online) Results of dimer mean-field theory for spins $S = \tfrac{1}{2}$ (left column) and $S = \tfrac{3}{2}$ (right column) on the 6H-perovskite dimer lattice, LT mode~1.
(a,b) Maps of the squared total spin $\langle \mathbf{S}_\mathrm{tot}^2 \rangle$ of the ground state across the $(J_2, J_3)$ parameter space, with a well-defined singlet region in both and the onset of mixing with excited states.
(c,d) Energies (scaled by $J_1$) of the dimer eigenstates as a function of $J_2/J_1$, for fixed $J_3/J_1 = 0.05$ (dashed lines in panels a,b).
(e,f) Corresponding $\langle \mathbf{S}_\mathrm{tot}^2 \rangle$ values for these eigenstates, illustrating the phase transition from a singlet and mixing between different spin multiplets.
(g,h) Composition of the ground-state wavefunction in the $\lvert S_z^{(2)}, S_z^{(3)} \rangle$ basis.
}
  \label{fig:dimer}
\end{figure}

Figure~\ref{fig:dimer} presents the results of the dimer mean-field theory for both $S = \tfrac{1}{2}$ and $S = \tfrac{3}{2}$.
The top row, panels (a) and (b), displays color maps of the squared total spin $\langle \mathbf{S}_\mathrm{tot}^2 \rangle$ across the $(J_2, J_3)$ parameter space.
In each case, a region of pure singlet ($\langle \mathbf{S}_\mathrm{tot}^2 \rangle = 0$) emerges for weak interdimer couplings $J_2$ and $J_3$.
Panels (c) and (d) show the energy spectrum of the dimer mean-field Hamiltonian as a function of $J_2/J_1$ for fixed $J_3/J_1 = 0.05$, marked by the dashed horizontal lines in panels (a) and (b).
This value of $J_3/J_1$ was selected because scanning $J_2/J_1$ reveals a quantum phase transition, visible as an inflection point in the energy spectra.
Panels (e) and (f) show the corresponding $\langle \mathbf{S}_\mathrm{tot}^2 \rangle$ values for each eigenstate along the same $J_2/J_1$ cut.
In the pure singlet phase, the ground state maintains $\langle \mathbf{S}_\mathrm{tot}^2 \rangle = 0$, but this value increases continuously above a critical $J_2/J_1$ as interdimer couplings drive hybridization with higher-spin sectors.
This behavior signals the onset of an interdimer-induced admixed phase, characterized by a finite local moment that supports the possibility of long-range dipolar magnetic order.
Finally, panels (g) and (h) display the ground-state wavefunction weights projected onto the product-state basis, labeled by the spin-$z$ components $S_z^{(2)}$ and $S_z^{(3)}$ of the two spins.
These panels illustrate how the singlet evolves continuously into a quantum superposition of multiplet components.
The overall phase of each eigenvector was fixed so all wavefunction amplitudes are real-valued.
For $S = \tfrac{1}{2}$, the transition involves redistribution to triplet states orthogonal to the singlet; for $S = \tfrac{3}{2}$, this redistribution also includes quintet and septet manifolds.

The $S = \tfrac{1}{2}$ case is presented as the minimal quantum model and serves to build intuition for the $S = \tfrac{3}{2}$ behavior, which is directly relevant for the $\mathrm{Ru}^{5+}$ ions in the compound Ba$_3$Zn$_{1-x}$Ca$_x$Ru$_2$O$_9$.
The extent of the singlet phase shrinks as the spin quantum number increases, consistent with the expectation that in the classical $S \rightarrow \infty$ limit, there is no singlet formation.
Within the finite-moment phase, there are both commensurate ($\mathbf{k}_0 = 0$) and incommensurate ($\mathbf{k}_0 = (h,h,0)$-type) solutions, as inherited from the LT analysis (see Appendix~\ref{sec:LT_appendix}). Wave vectors are reported in reciprocal-lattice units (r.l.u.) of the 6H-perovskite cell, $\mathbf{Q}=(h,k,l)=h\mathbf{a}^*+k\mathbf{b}^*+l\mathbf{c}^*$, with $|\mathbf{a}^*|=|\mathbf{b}^*|=2\pi/a$ and $|\mathbf{c}^*|=2\pi/c$.

\begin{figure}[th]
  \centering
  \includegraphics[width=\columnwidth]{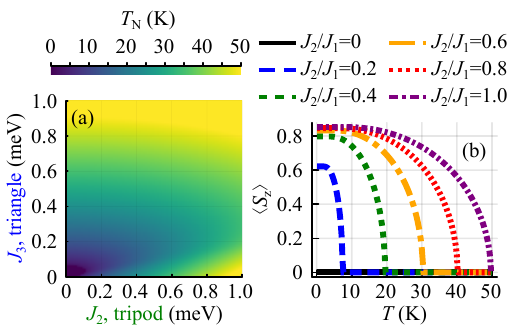}
  \caption{(Color online)
  Results of finite-temperature dimer mean-field theory for $S = \tfrac{3}{2}$ on the 6H-perovskite dimer lattice, for LT mode~1.
  (a) Mean-field ordering temperature $T_\mathrm{N}$ in kelvin for $J_1 = 1~\mathrm{meV}$, as a function of interdimer couplings $J_2$ and $J_3$.
  (b) Temperature dependence of the staggered moment $\langle S_z \rangle$ for selected $J_2$ values at fixed $J_3 = 0.05$.}
  \label{fig:dimer-T}
\end{figure}

\begin{figure*}[!ht]
  \centering
  \includegraphics[width=\textwidth]{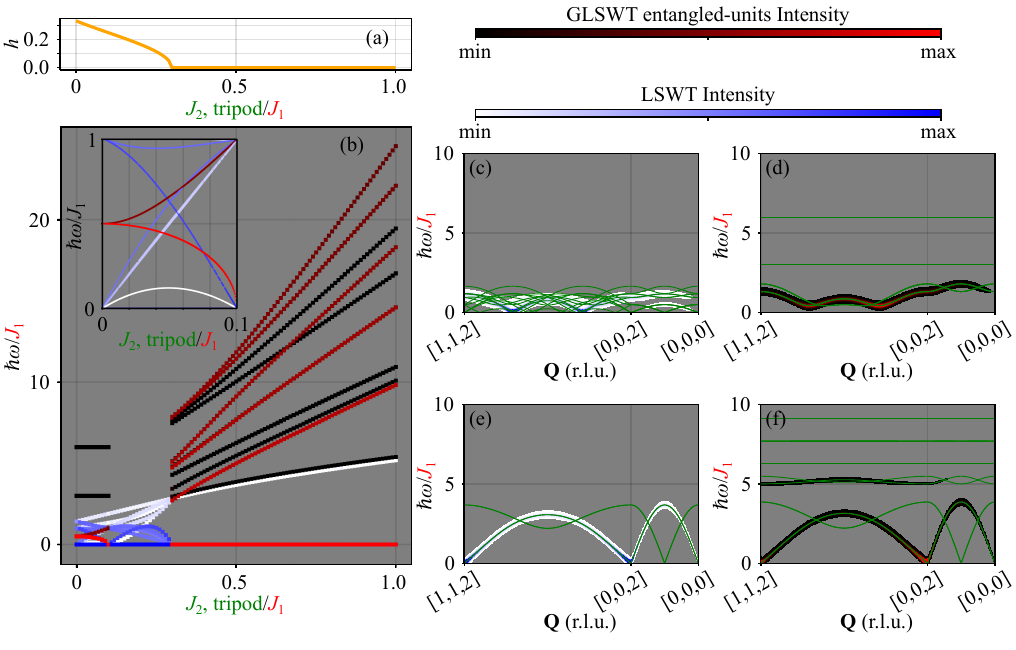}
  \caption{(Color online)
  LSWT of the 6H-perovskite dimer lattice with varying interdimer exchange $J_2/J_1$ (with $J_3/J_1 = 0.05$, $J_4 = 0$) at zero temperature, shown for both LSWT and entangled-unit GLSWT calculations.
  (a) Ordering wavevector component $h$ of $\mathbf{k}_0 = (h,h,0)$ as a function of $J_2/J_1$.
  (b) Comparison of excitation spectra at $\mathbf{k}_0$ computed using traditional LSWT (white to blue colormap) and GLSWT entangled-unit formalism (black to red colormap).
  The inset magnifies the low-$J_2$ regime, where the entangled-unit model yields a pure singlet ground state.
  Intensities in (b) are plotted on a $\log_{10}$ scale.
  Panels (c) and (e) show the intensity-clipped neutron dynamic structure factor $S(\mathbf{Q}, \hbar \omega)$ for representative values $J_2/J_1 = 0.05$ and $J_2/J_1 = 0.5$, respectively, with overlaid magnon modes (green).
  Panels (d) and (f) show the corresponding results from the entangled-unit approach.
  }
  \label{fig:LSWT}
\end{figure*}

To determine the onset of long-range magnetic order, we extend the dimer mean-field theory to finite temperatures.
In this formulation, the mean-field Hamiltonian remains unchanged, but the spin expectation values are computed using thermally weighted averages over all eigenstates of the dimer:
\begin{equation}
\langle \mathbf{S} \rangle = \frac{\sum_i \langle \psi_i | \mathbf{S} | \psi_i \rangle\, e^{-E_i / k_B T}}{\sum_i e^{-E_i / k_B T}},
\end{equation}
where $\psi_i$ and $E_i$ are the eigenvectors and eigenvalues of the self-consistent dimer Hamiltonian at a given temperature $T$.
Temperatures are given in kelvin, and we take $J_1 = 1~\mathrm{meV}$ as the energy scale.
This approach yields the mean-field transition temperature $T_\mathrm{N}$ as a function of $J_2$ and $J_3$, shown in Fig.~\ref{fig:dimer-T}(a) for the $S = \tfrac{3}{2}$ case.
Figure~\ref{fig:dimer-T}(b) presents the temperature dependence of the staggered moment $\langle S_z \rangle$ for fixed $J_3 = 0.05$ meV and several representative values of $J_2$.
As temperature decreases, the moment rises continuously from zero at $T_\mathrm{N}$, reflecting the second-order character of the mean-field transition.
At low temperature, the moment saturates to the ground-state maximum (which is less than the full $S = \tfrac{3}{2}$ value), and its magnitude increases with $J_2$, consistent with stronger interdimer coupling favoring moment formation.

\subsection{\label{sec:LSWT}Linear Spin-wave Theory}

To better understand the collective excitations of the 6H-perovskite dimer lattice, we employed both traditional LSWT as well as a generalized LSWT (GLSWT) using an ``entangled-unit'' formalism ~\cite{Dahlbom2024} as implemented in \texttt{Sunny.jl}~\cite{Dahlbom2025}.
The GLSWT approach captures the full 16-dimensional local Hilbert associated with the $S=\tfrac{3}{2}$ dimer on each $J_1$-bond and enables the direct modeling of multiplet mixing. 
We consider a representative exchange parameter set along the same line cut as the dimer mean-field calculations, $J_3/J_1 = 0.05$, and $J_4 = 0$, varying $J_2/J_1$ to explore the evolution of magnetic excitations.
For the LSWT calculations, we use the single-$\mathbf{k}$ spiral formalism~\cite{Toth2015} when the system becomes incommensurate, and we verify that the predicted ground-state wavevector from LT analysis agrees with direct energy minimization within \texttt{Sunny.jl}.
As shown in Fig.~5(a), the system smoothly transitions from $\mathbf{k} = \left( \tfrac{1}{3}, \tfrac{1}{3}, 0 \right)$ at $J_2 = 0$ to incommensurate order, and then to commensurate $\mathbf{k} = (0, 0, 0)$ order as $J_2/J_1$ increases.

Figure~\ref{fig:LSWT}(b) compares the spin-wave spectra at the LT wavevector $\mathbf{k}_0$ using both methods.
At $J_2=0$ (decoupled dimer layers and weak within-layer coupling), the GLSWT calculation yields a gapped, nonmagnetic singlet ground state with dispersive triplons that exhibit a minima at $\mathbf{k}_0$, while the traditional LSWT calculation yields a gapless Goldstone mode due to spontaneous symmetry breaking. Representative spectra at a slightly larger value, $J_2/J_1 = 0.05$, [Figs.~\ref{fig:LSWT}(c,d)] show broadly similar dispersions across momentum space for both approaches, though the entangled-unit model remains gapped, exhibits a slightly higher-energy mode, and includes an additional excitation branch. As $J_2$ grows, the triplon band in the GLSWT calculation softens, eventually becoming gapless with the emergence of magnetic order. 
In the regime with finite dipolar order and incommensurate wavevector, the spin-wave spectra are not computed for the GLSWT model due to the computational cost of large supercells.
 At $J_2/J_1 = 0.5$ [Figs.~\ref{fig:LSWT}(e,f)], the two methods converge, yielding nearly indistinguishable spectra aside from minor scaling in mode energies and residual multiplet mode intensity, supporting the robustness of the LSWT  description in the large-$J_2$ regime.

Importantly, the entangled-units formalism captures the stabilization of the pure singlet phase in the low-$J_2$ limit, where all higher multiplet components remain unpopulated, as well as the transition to a finite dipolar order with reduced on-site magnetization due to mixing between singlet and higher-spin multiplets.
Notably, within the GLSWT description, the phase transition from the singlet to the interdimer-induced admixed phase is accompanied by the closing of the excitation gap. We note that the entangled-units calculation is most accurate for smaller values of $J_2/J_1$. Moreover, as a general rule, semi-classical methods such as the entangled-units formalism tend to overestimate the stability of ordered phases near quantum phase transitions, meaning that the the value of $J_2/J_1$ at which the triplon gap closes and magnetic order emerges is likely larger than what is predicted here (see \cite{Zhang2025} and references therein). One can therefore expect that the portion of the phase diagram corresponding to a pure singlet ground state (no magnetic moment) is in fact larger than what is reported here. 

\begin{figure}[htbp]
  \centering
  \includegraphics[width=\columnwidth]{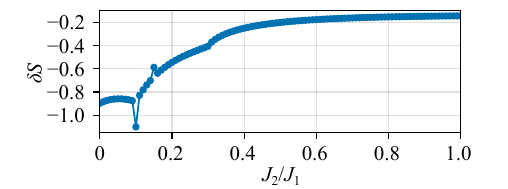}
  \caption{(Color online)
  LSWT zero-point spin correction $\delta S$ for the 6H-perovskite dimer lattice as a function of $J_2$, with fixed $J_1 = 1$, $J_3 = 0.05$, and $J_4 = 0$.}
  \label{fig:dS}
\end{figure}

Within the traditional LSWT formalism, we calculated the quantum correction $\delta S$ to the ordered moment, which provides a measure of the extent to which zero-point fluctuations reduce magnetic order.
Fig.~\ref{fig:dS} shows $\delta S$ for a range of $J_2/J_1$ values with fixed $J_3/J_1 = 0.05$ and $J_4 = 0$.
When $J_2/J_1 \leq 0.1$, this analysis suggests a very strong suppression of the magnetic moment. 
The entangled-unit GLSWT analysis indicates that the ground state in this region of the phase diagram is product of pure singlets, which cannot be represented in the traditional (large-$S$) semiclassical framework. The large $\delta S$ values are therefore expected, but it is not possible to tell from this analysis whether they may be attributed entirely to the fluctuations within localized dimers or to some collective effect. $J_2/J_1 \approx 0.1$ corresponds to the emergence of a finite magnetic moment in the entangled-units formalism and the onset of incommenensurate order; as expected, the strongest quantum fluctuations outside of the singlet region of the phase diagram occur in this incommensurate phase, with a correspondingly large suppression of the ordered moment.
% the classical ground state used in the LSWT formalism is ordered, but the corresponding GLSWT calculation remains in a singlet phase with no dipolar order.
% Beyond this threshold, the entangled-unit model develops a finite moment, and the traditional LSWT description becomes qualitatively valid for describing both the magnetic excitations and the order parameter.
A clear kink appears in $\delta S$ near $J_2/J_1 \approx 0.25$, corresponding to the crossover from incommensurate to commensurate ($\mathbf{k}_0 = 0$) magnetic order.
Given the close agreement between the LSWT and GLSWT spectra outside the singlet phase (larger $J_2/J_1$), the LSWT values of $\delta S$ serve as a useful estimate for moment reduction in the full quantum system.
% While these results are at zero temperature, finite-temperature magnon populations would further reduce the ordered moment.

To summarize, LSWT calculations on the 6H-perovskite dimer lattice reveal complementary insights from traditional LSWT and GLSWT entangled-unit approaches.
The entangled-unit formalism captures the stabilization of a gapped singlet phase in the weakly interacting limit, while LSWT model effectively describes the ordered regime and its excitations.
Both methods exhibit similar spectra once dipolar order emerges, supporting the use of LSWT as a reliable low-energy theory when well outside of the singlet dome.
Additionally, LSWT corrections to the ordered moment and zero-point energies highlight the role of quantum fluctuations in selecting ordering vectors and suppressing magnetic order.

\section{\label{sec:BZCRO}Connection to real materials}

\subsection{\label{sec:DFT}Density Functional Theory}

Having established the phase behavior of the antiferromagnetically coupled 6H-perovskite dimer lattice through model Hamiltonian analysis, we now turn to first-principles calculations to connect this framework to Ba$_3$Zn$_{1-x}$Ca$_x$Ru$_2$O$_9$.
To this end, we employ DFT calculations for select systems to extract exchange parameters by fitting Heisenberg models of the form
\begin{equation}
H = \sum_{\langle ij \rangle} J_{ij} \sigma_i \sigma_j S^2,
\end{equation}
to the total energies of different spin configurations, where \( \sigma_i = \pm 1 \) encodes the spin direction (up or down) at site \( i \) in a collinear configuration.
The model systems studied include BZRO, BCRO, and a magnetically dilute compound Ba$_3$Zn(Ru$_{1-x}$Sb$_x$)$_2$O$_9$ (BZRSO), in which magnetic Ru$^{5+}$ ions reside on a background of diamagnetic Sb$^{5+}$.

A detailed account of the DFT methodology, spin configuration labeling, energy mapping, variation of the effective Hubbard parameter $U_\mathrm{eff}$, and exchange extraction is provided in the Appendix~\ref{sec:DFT_appendix}.
The first key finding is that hybridization between Ru and the surrounding ligands reduces the magnetic moment, consistent with earlier DFT results for Ba$_3$CoRu$_2$O$_9$~\cite{Streltsov2013}.
For both BZRO and BCRO, we find $\mu_\mathrm{Ru} = 1.9~\mu_\mathrm{B}$, reduced from the full-spin value of $3~\mu_\mathrm{B}$.
Single-ion anisotropy terms of the form $D S_z^2$ are found to favor easy-axis alignment along the crystallographic $c$-axis, with $D = -0.03~\mathrm{meV}$ for BZRO and $D = -0.16~\mathrm{meV}$ for an isolated monomer in BZRSO.

\begin{table}[hbt]
  \caption{DFT-derived exchange parameters (in meV) at
    $U_{\mathrm{eff}}=2.3\,\mathrm{eV}$. Ratios are normalized to $J_1$.}
  \label{tab:DFT}
  \centering
  \begin{ruledtabular}
    \begin{tabular}{*{8}{c}}
      System &  $J_1$ & $J_2$  & $J_3$  & $J_4$  & $J_2/J_1$ & $J_3/J_1$ & $J_4/J_1$ \\
      \hline
      BZRSO & 19.10 & 7.39   & 1.35   & 0.30   & 0.387     & 0.0706    & 0.0155 \\
      BCRO  & 30.71 & 5.85   & 1.55   & 0.26   & 0.191     & 0.0504    & 0.0085  \\
      BZRO  & 19.33 & 6.93   & 1.19   & 0.27   & 0.358     & 0.0615    & 0.0142  \\
    \end{tabular}
  \end{ruledtabular}
\end{table}

The normalized exchange constants at $U_\mathrm{eff} = 2.3~\mathrm{eV}$ provide a clear comparative framework for understanding the impact of B-site substitution (Zn$^{2+} \rightarrow$ Ca$^{2+}$) on magnetic interactions (see Table~\ref{tab:DFT}).
Substituting the smaller Zn$^{2+}$ ion with the larger Ca$^{2+}$ weakens interdimer superexchange and strengthens intradimer coupling.
BZRSO exhibits slightly enhanced interdimer exchanges compared to BZRO and slightly weaker intradimer coupling, but overall the trends are similar—suggesting that BZRSO may serve as a reliable proxy for BZRO in theoretical analysis.

\subsection{\label{sec:dilute}Neutron spectroscopy of
  Ba$_3$Zn(Ru$_{1-x}$Sb$_x$)$_2$O$_9$}

\begin{figure*}[!htb]
  \centering
  \includegraphics[width=\textwidth]{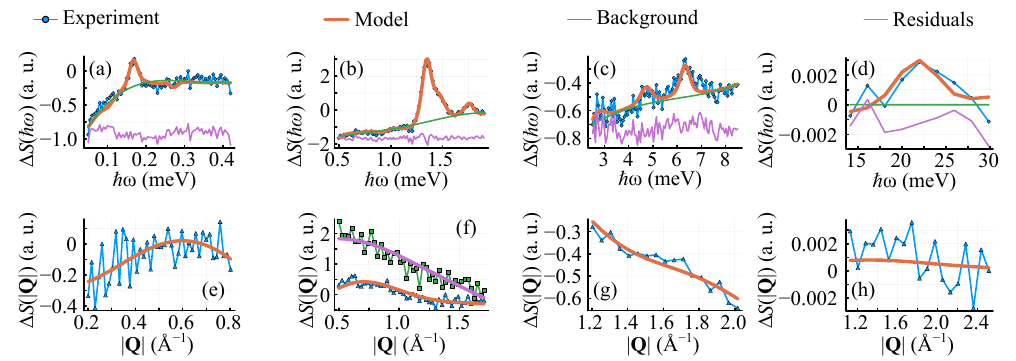}
  \caption{(Color online) One-dimensional cuts from the neutron spectroscopy data on BZRSO, showing the temperature-difference signal $\Delta S = S(T_{base}) - S(300~\mathrm{K})$.
Panels (a)–(d) show energy-dependent cuts at fixed momentum transfer $|\mathbf{Q}|$ for the four incident energies used:
(a) $E_i = 1.00$ meV with $|\mathbf{Q}| = [0.2, 0.8]$ Å$^{-1}$,
(b) $E_i = 3.32$ meV with $|\mathbf{Q}| = [0.5, 1.7]$ Å$^{-1}$,
(c) $E_i = 12$ meV with $|\mathbf{Q}| = [1.2, 2.0]$ Å$^{-1}$, and
(d) $E_i = 100$ meV with $|\mathbf{Q}| = [1.1, 2.0]$ Å$^{-1}$.
Curves are offset vertically for clarity.
Panels (e)–(h) show momentum-dependent cuts at fixed energy transfer $\hbar\omega$ for the same $E_i$:
(e) $E_i = 1.00$ meV with $\hbar\omega = [0.14, 0.20]$ meV,
(f) $E_i = 3.32$ meV with $\hbar\omega = [1.2, 1.4]$ meV (upper curve) and $[1.62, 1.84]$ meV (lower curve),
(g) $E_i = 12$ meV with $\hbar\omega = [4.0, 7.5]$ meV, and
(h) $E_i = 100$ meV with $\hbar\omega = [20, 28]$ meV.
In the $|\mathbf{Q}|$-cuts, experimental data are shown as points connected by lines, and model predictions are overlaid as continuous curves. Hamiltonian parameters for the model are in Table~\ref{tab:INSparams}.}
  \label{fig:INS-1d}
\end{figure*}

To build upon the insights gained from DFT calculations, we now turn to INS measurements of BZRSO. The use of INS on magnetically dilute compounds to probe the exchange interactions of a related dense system has precedent, notably in LaMn$_{0.1}$Ga$_{0.9}$O$_3$, which highlights both the strengths and subtleties of this approach~\cite{Furrer2011}. In BZRSO, the synthesis target was for $5\%$ of the B-site ions to be Ru$^{5+}$ (4$d^3$, $S = \tfrac{3}{2}$), and the remaining $95\%$ to be diamagnetic Sb$^{5+}$ (4$d^{10}$), with respective ionic radii of 56.5 pm and 60.0 pm~\cite{Shannon1976}. The resulting material consists of isolated monomers, dimers, trimers, and larger clusters of Ru$^{5+}$ ions embedded in a nonmagnetic matrix.

Within this chemically dilute environment, magnetic excitations can be modeled via exact diagonalization (ED) of isolated clusters. This approach enables extraction of the underlying exchange interactions without requiring a full many-body treatment as would be necessary for the dense BZRO system. While the exchange constants obtained from BZRSO clusters may not map directly onto those of BZRO, comparisons with cluster-based DFT calculations provide useful context for interpreting trends. Synthesis and measurement details, cluster distribution estimates, $|\mathbf{Q}|$ vs. $\hbar\omega$ intensity maps, and the structure factor equation are provided in the Appendix~\ref{sec:dilute_appendix}.

Quantitative fitting of the neutron spectra was performed by integrating over momentum to extract intensity as a function of energy transfer, $\hbar\omega$, and comparing with momentum-dependent scans using independent background fits, as shown in Fig.~\ref{fig:INS-1d}. The fitting procedure began with the $E_i = 3.32$\,meV dataset, which showed distinct peaks from monomers (\(\approx 1.3\,\text{meV}\)), $J_3$-dimers (\(\approx 1.75\,\text{meV}\)), and $J_4$-dimers (\(\approx 1.45\,\text{meV}\)), Fig.~\ref{fig:INS-1d}(b). Cluster populations were fixed assuming a 2.5\% Ru concentration, as inferred from relative peak intensities. This differs from the nominal 5\% synthesis target, likely due to unreacted material or intensity variations in $\hbar\omega$. However, since peak positions---not absolute intensities---are the primary constraint, this uncertainty has limited impact on the extracted parameters.

A spline background was co-refined alongside the model by fitting energy regions devoid of magnetic excitations. This background accounts for contributions from non-magnetic scattering, phonons, and unresolved clusters. Momentum scans for $E_i = 3.32$\,meV are shown in Fig.~\ref{fig:INS-1d}(f), comparing a monomer-dominated region and one containing $J_3$-dimers. The monomer signal decays monotonically with increasing $|\mathbf{Q}|$, while the dimer exhibits interference oscillations from finite inter-ion separation. Independent background fits included constant and linear terms. Due to this, deviations from the assumed magnetic form factor may be partially absorbed by the background and are not explicitly refined.

The single-ion anisotropy $D$ was extracted from this dataset and held fixed in subsequent fits. The $E_i = 1.00$\,meV data [Fig.~\ref{fig:INS-1d}(a)] show a single sharp excitation at \(\approx 0.16\,\text{meV}\) attributed to a $J_3$-dimer transition. No additional parameters were refined beyond those obtained from the 3.32\,meV fit. The corresponding momentum scan [Fig.~\ref{fig:INS-1d}(e)] reveals oscillations consistent with the $J_3$-dimer structure factor.

The $E_i = 12.0$\,meV data [Fig.~\ref{fig:INS-1d}(c)] show a doublet from a $J_2$-dimer, split by the anisotropy $D$, allowing extraction of $J_2$. The associated momentum dependence [Fig.~\ref{fig:INS-1d}(g)] shows features consistent with the expected structure factor. The most challenging dataset was $E_i = 100$\,meV [Fig.~\ref{fig:INS-1d}(d)], where the $J_1$-dimer excitation occurs at low-$|\mathbf{Q}|$ and high energy transfer---conditions with reduced flux, broader resolution, and increased phonon background. Additionally, $J_1$-dimers are statistically less common. Still, a weak feature appears near the expected energy. The corresponding $|\mathbf{Q}|$-cut [Fig.~\ref{fig:INS-1d}(h)] is noisy and largely inconclusive.

\begingroup
%\squeezetable
\begin{table}[!bht]
  \caption{Best-fit exchange parameters and single‑ion anisotropy $D$ (in meV) from ED modeling of INS data on dilute BZRSO. Ratios normalized to $J_{1}$.}
  \label{tab:INSparams}
  \centering
  \begin{ruledtabular}
  \begin{tabular}{ *{8}{c} }
    $D$ & $J_1$ & $J_2$ & $J_3$ & $J_4$ & $J_2/J_1$ & $J_3/J_1$ & $J_4/J_1$ \\
    \hline
    -0.67 & 21.7 & 5.65 & 0.66 & 0.085 & 0.260 & 0.030 & 0.004
  \end{tabular}
  \end{ruledtabular}
\end{table}
\endgroup

These INS measurements on dilute BZRSO provide a valuable benchmark for the exchange hierarchy and anisotropy predicted by DFT for the parent compound BZRO. The extracted exchange ratios align well with DFT results at $U_\mathrm{eff} = 2.3$\,eV. The data confirm that $J_1$ and $J_2$ dominate the interaction network, consistent with the view that BZRO is inherently three-dimensional---rather than quasi-2D, which would require dominant $J_3$.
The measured anisotropy $D$ is easy-axis and larger in magnitude than DFT predictions for the dense BZRO system. This enhancement is expected in dilute systems, where reduced screening and local symmetry breaking strengthen crystal-field effects. By contrast, the full BZRO lattice retains higher symmetry and stronger hybridization, both of which suppress anisotropy. Thus, the large $D$ observed in BZRSO supports the conclusion that single-ion anisotropy in dense BZRO is small---an important detail for modeling its collective behavior.
Additionally, these results indicate that DFT tends to overestimate interdimer exchanges. In a dense magnetic lattice, further renormalization from quantum and thermal fluctuations is expected. Finally, the consistently weak $J_4$ found in both DFT and experiment supports its omission from minimal models of BZRO.

\subsection{\label{sec:darriet}Linear Spin-wave Theory Modeling Previously Reported
  Ba$_3$CaRu$_2$O$_9$ Neutron
  Spectroscopy}

\begin{figure*}[!bht]
  \centering
  \includegraphics[width=\linewidth]{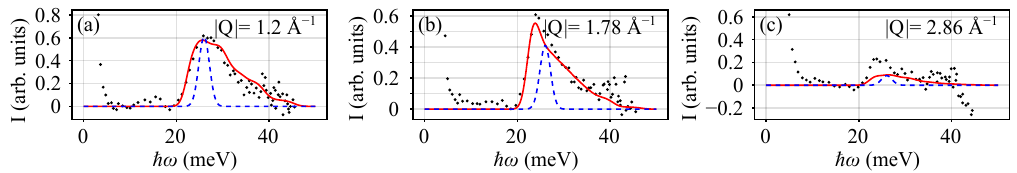}
  \caption{(Color online)
  Comparison between digitized neutron inelastic scattering data (black points), \texttt{Sunny.jl} entangled-unit model calculations (red solid curves), and a non-interacting $J_1$-only dimer model (blue dashed curves) for BCRO at three momentum transfers: (a) $|\mathbf{Q}| = 1.2$\,\AA$^{-1}$, (b) $|\mathbf{Q}| = 1.78$\,\AA$^{-1}$, and (c) $|\mathbf{Q}| = 2.86$\,\AA$^{-1}$. Data are from Ref.~\cite{Darriet1983} and represent the difference between 6\,K and 300\,K measurements after Bose-factor correction. Model calculations use the best-fit parameters from entangled-units linear spin-wave theory described in the text and summarized in Table~\ref{tab:BCRO_fit}.}
  \label{fig:BCROfits}
\end{figure*}

In this section, we revisit the published INS data on BCRO~\cite{Darriet1983}. In that work, neutron spectra were measured on a powder sample at three momentum transfers---$|\mathbf{Q}| = 1.2$, 1.78, and 2.86\,\AA$^{-1}$---at temperatures of 6\,K and 300\,K. While the original analysis interpreted the data using an isolated dimer model, the approach neglected weak but crucial interdimer interactions. We reanalyzed these data using the previously discussed $J_1$, $J_2$, and $J_3$ exchange parameters in the Hamiltonian, with $J_4 = 0$ and $D = 0$ to simplify the model as motivated by the DFT and BZRSO neutron experiments. Additional details are provided in the Appendix~\ref{sec:darriet_appendix}.

We digitized the published spectra and performed new fits using the \texttt{Sunny.jl} entangled-units formalism to capture both intradimer and interdimer couplings. To isolate the magnetic signal, we treated the 300\,K data as a non-magnetic background and subtracted it from the 6\,K spectra after applying Bose factor corrections. Though this method captures much of the non-magnetic scattering, residuals such as a peak near 15\,meV---likely of phonon origin---persist in the subtracted data. Discrepancies in model intensity, especially at higher momentum transfer ($|\mathbf{Q}| = 2.86$\,\AA$^{-1}$), may also reflect limitations in the background subtraction or Ru$^{5+}$ form factor modeling, for which we used the Ru$^{1+}$ tabulation.

Figure~\ref{fig:BCROfits} presents the resulting fits compared to the experimental spectra and an isolated dimer model ($J_1=26$ meV) at each $|\mathbf{Q}|$. Optimized parameters for the entangled-units fit are shown in Table~\ref{tab:BCRO_fit}, and the non-interacting $J_1$-only dimer model was not optimized. While the isolated dimer model yields a single non-dispersive peak at the singlet-triplet transition energy, the interacting dimer model captures the broader structure and subtle momentum dependence more accurately, such as the decreasing mode position with increasing $|\mathbf{Q}|$ reported in the original paper. Despite the small magnitude of interdimer exchanges, their inclusion significantly reshapes the excitation spectrum, confirming the importance of treating BCRO as an interacting dimer system.

\begin{table}[htbp] % floats at here/top/bottom as needed
  \caption{Neutron spectroscopy–derived exchange constants (in meV) for BCRO.}
  \label{tab:BCRO_fit}
  \centering
  \begin{ruledtabular}
    \begin{tabular}{cccccc}
      $J_1$                 & $J_2$                  & $J_3$                  & $J_2/J_1$ & $J_3/J_1$ \\
      \hline
      $30.6\,\pm\,0.3$      & $1.57\,\pm\,0.08$      & $0.60\,\pm\,0.04$      & $0.051$   & $0.020$
    \end{tabular}
  \end{ruledtabular}
\end{table}

The extracted parameters place BCRO well within the singlet region of the phase diagram, with no long-range dipolar order, consistent with its non-magnetic ground state. Nevertheless, the calculated spectrum exhibits clear triplon dispersion and spectral broadening beyond instrumental resolution, emphasizing the nontrivial role of interdimer coupling even in the absence of magnetic order.

\subsection{\label{sec:terasaki}Dimer Mean-field Theory Modeling of Previously Reported
  Ba$_3$Zn$_{1-x}$Ca$_x$Ru$_2$O$_9$
  Magnetization Data}

\begin{figure}[htbp]
  \centering
  \includegraphics[width=\columnwidth]{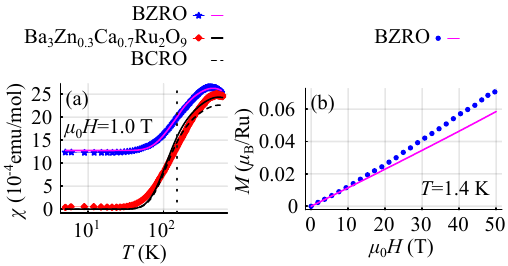}
  \caption{(Color online)
  Comparison of dimer mean-field model fits to 6H-perovskite dimer lattice experimental data digitized from Ref.~\cite{Terasaki2017} 
  (a) DC magnetic susceptibility versus temperature. The vertical dashed line at $T = 150$\,K shows the mean-field ordering temperature of the BZRO model. 
  (b) High-field magnetization. Experimental data are symbols. Solid lines represent dimer mean-field model fits using $S = \tfrac{3}{2}$ with best-fit exchange parameters described in the text and summarized in Table~\ref{tab:Terasaki_fit_params}.}
  \label{fig:magfit}
\end{figure}

Experimental data on the magnetization of Ba$_3$Zn$_{1-x}$Ca$_x$Ru$_2$O$_9$ have been reported~\cite{Terasaki2017}, including temperature-dependent DC magnetic susceptibility for BZRO and Ba$_3$Zn$_{0.7}$Ca$_{0.3}$Ru$_2$O$_9$, as well as pulsed-field magnetization up to 50\,T for BZRO. A subset of these data was digitized and analyzed using the finite-temperature dimer mean-field theory for $S = \tfrac{3}{2}$, with exchanges $J_1$, $J_2$, and $J_3$. The resulting fits are shown in Fig.~\ref{fig:magfit}, the best-fit parameters are summarized in Table~\ref{tab:Terasaki_fit_params}, and loss function maps are shown in Appendix~\ref{sec:terasaki_appendix}.

\begin{table}[hbt]
  \caption{Magnetic susceptibility–derived exchange constants (in~meV) for BZRO and Ba$_3$Zn$_{0.7}$Ca$_{0.3}$Ru$_2$O$_9$. Ratios normalized to $J_1$.}
  \label{tab:Terasaki_fit_params}
  \centering
  \begin{ruledtabular}
    \begin{tabular}{lcccccc}
      Compound &
      $J_1$ & $J_2$ & $J_3$ &
      $J_2/J_1$ & $J_3/J_1$ \\
      \hline
      BZRO       & 25.2 & 3.16 & 0.31 & 0.125 & 0.012 \\
      Ba$_3$Zn$_{0.7}$Ca$_{0.3}$Ru$_2$O$_9$ & 30.0 & 0.43 & 0.16 & 0.014 & 0.005 \\
    \end{tabular}
  \end{ruledtabular}
\end{table}

For BZRO and Ba$_3$Zn$_{0.7}$Ca$_{0.3}$Ru$_2$O$_9$, the susceptibility was fit over the full temperature range using the thermally averaged local moment computed self-consistently from the dimer mean-field Hamiltonian. For BCRO, the parameters obtained from our neutron scattering analysis of Darriet \emph{et al.}~\cite{Darriet1983} were used to compute the magnetic susceptibility. In both BCRO and Ba$_3$Zn$_{0.7}$Ca$_{0.3}$Ru$_2$O$_9$, the ground state is a non-magnetic singlet, and the susceptibility increases with temperature due to thermal excitation of higher-spin multiplets. In contrast, the dimer mean-field solution for BZRO yields an interdimer-induced admixed ground state with a finite dipolar moment of $0.57\,\mu_\mathrm{B}$ and a predicted onset of mean-field order at $T = 150$\,K. This interdimer-induced admixed state explains the small but nonzero susceptibility observed at low temperatures.

The high-field magnetization for BZRO, computed using the susceptibility-fit parameters, matches the experimental data well at low fields. Above 10\,T, however, the model slightly underestimates the measured magnetization. This deviation may reflect the absence of short-range or dynamic correlations in the mean-field treatment, which could then be partially suppressed at high fields, or this deviation may be due to the presence of magnetic impurities. For BCRO and Ba$_3$Zn$_{0.7}$Ca$_{0.3}$Ru$_2$O$_9$, the dimer models remain in a singlet phase across this field range, though sufficiently high fields would eventually mix in magnetic multiplets and induce a finite moment.

While these exchange parameters are specific to the applied dimer mean-field framework, complementary insights can be gained by evaluating the traditional LSWT correction to the ordered moment for BZRO. For the dimer mean-field parameter set, this correction is $\delta S = -0.42$ ($-0.84\,\mu_\mathrm{B}$), which exceeds the mean-field moment itself, highlighting the strong role of quantum fluctuations. Furthermore, the presence of low-lying multiplets near the mean-field ordering temperature suggests that specific heat measurements in this temperature range would be dominated by thermal (de)population of these excited states. Altogether, the dimer mean-field theory provides a minimal yet effective description that captures the trends in susceptibility across the Ba$_3$Zn$_{1-x}$Ca$_x$Ru$_2$O$_9$ series, including the suppression of moment and emergence of singlet behavior with increasing Ca substitution.

\section{\label{sec:summary}Summary and Conclusions}

%In this work, w
We developed a description of the 6H-perovskite dimer lattice with antiferromagnetic interactions, combining classical LT theory, quantum dimer mean-field modeling, traditional LSWT, and entangled-unit GLSWT. The substitution series Ba$_3$Zn$_{1-x}$Ca$_x$Ru$_2$O$_9$ is an experimental realization of these phases and contains previously unquantified magnetic behavior for small $x$. To understand the magnetic energies in these compounds, DFT calculations were performed and fit to Heisenberg Hamiltonians, and INS measurements on magnetically dilute BZRSO were modeled using ED of clusters. An existing report on neutron spectroscopy of BCRO was digitized and analyzed with GLSWT to extract exchange energies and quantify the magnetic ground state. Similarly, magnetization data for BZRO and Ba$_3$Zn$_{0.7}$Ca$_{0.3}$Ru$_2$O$_9$ were digitized and analyzed using dimer mean-field theory. These analyses show that $J_1 > J_2 > J_3 > J_4$, and that the triangular lattice interaction does not dominate the physics for Ba$_3$Zn$_{1-x}$Ca$_x$Ru$_2$O$_9$, but that the interdimer interactions are highly three-dimensional. As such, while the anomalous magnetism of BZRO was previously suggested to be potentially quadrupolar or nematic due to the triangular motif of the dimers~\cite{Tanaka2020}, that model is not consistent with our finding that $J_2 > J_3$ in BZRO.

\begin{figure}[thbp]
  \centering
  \includegraphics[width=\columnwidth]{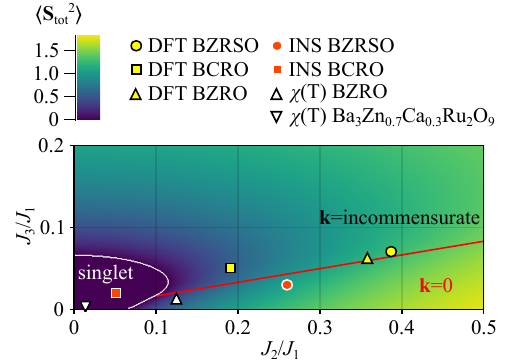}
  \caption{(Color online) 
Phase diagram for the 6H-perovskite dimer lattice with $S=\tfrac{3}{2}$. The color map shows the squared total spin $\langle \mathbf{S}_\mathrm{tot}^2 \rangle$ from dimer mean-field theory, indicating the transition from a non-magnetic singlet ($\langle \mathbf{S}_\mathrm{tot}^2 \rangle = 0$) to a finite-moment ground state as a function of normalized interdimer couplings $J_2/J_1$ and $J_3/J_1$. The red line denotes the LT transition between commensurate ($\mathbf{k} = 0$) and incommensurate ordering wavevectors.  The white line outlines the region where a singlet ground state is stabilized by mean-field theory. Estimates of exchange ratios from DFT, INS, and magnetic susceptibility [$\chi(T)$] modeling are overlaid as labeled symbols for several compounds.
}

  \label{fig:summary}
\end{figure}

To visualize these results, we present in Fig.~\ref{fig:summary} a schematic overview of a phase diagram and results from the preceding Sections for the Ba$_3$Zn$_{1-x}$Ca$_x$Ru$_2$O$_9$ family. The color map shows the dimer mean-field $\langle \mathbf{S}_\mathrm{tot}^2 \rangle$ as function of normalized interdimer exchange couplings $J_2/J_1$ and $J_3/J_1$, with phase boundaries between the singlet, the $\mathbf{k} = 0$ ordered phase, and the incommensurate ordered phase. Symbols are overlaid for the best-fit parameters from DFT, neutron spectroscopy, and magnetization modeling. These results show that both BCRO and Ba$_3$Zn$_{0.7}$Ca$_{0.3}$Ru$_2$O$_9$ are found experimentally to have singlet ground states, while BZRO and BZRSO are experimentally found to have $\mathbf{k} = 0$ ordered ground states in regions with decreased dipolar moments. The DFT results misplace BCRO to be outside the singlet region, but do reproduce the experimental trends of relative placements for BCRO, BZRO, and BZRSO.

A central conclusion of this study is that magnetic moment suppression arises in Ba$_3$Zn$_{1-x}$Ca$_x$Ru$_2$O$_9$ through three distinct mechanisms:

\begin{enumerate}
    \item Quantum fluctuations inherent to low-dimensional and frustrated antiferromagnets, captured by linear spin-wave theory as a correction to the ordered moment ($\delta S = -0.4$ for BZRO using the parameters derived from dimer mean-field theory susceptibility fits);
    \item Hybridization with surrounding oxygen ligands, which decreases the local Ru$^{5+}$ moment, as revealed by DFT (1.9\,$\mu_\mathrm{B}$ for BZRO, down from the full 3\,$\mu_\mathrm{B}$); and
    \item Singlet--multiplet mixing induced by interdimer exchange, leading to admixed ground states above a critical interdimer interaction strength, with finite but reduced dipolar moments in the dimer mean-field theory (0.57\,$\mu_\mathrm{B}$ down from the full 3\,$\mu_\mathrm{B}$ for BZRO using the parameters derived from dimer mean-field theory susceptibility fits).
\end{enumerate}

Taken together, these mechanisms can conspire to suppress long-range dipolar magnetic order in BZRO below the detection limits of the reported neutron diffraction methods---or even eliminate it entirely. The $T = 4.2$\,K Mössbauer signal that shows a clear internal field for BZRO may signify either short-range correlations or long-range order with a small dipolar moment~\cite{Fernandez1980}. The potential for minor structural disorder to disrupt a weakly stabilized ordered state is also important to consider. Additionally, this framework provides a coherent explanation for the evolution of magnetic properties with Zn-to-Ca substitution. Overall, this description connects microscopic exchange interactions to macroscopic behavior and offers a roadmap for understanding moment suppression and anomalous magnetism in a broader class of quantum dimer materials.

Several avenues remain for advancing the understanding of quantum magnetism in the 6H-perovskite dimer lattice. Our analysis suggests that BZRO in particular lies intriguingly close to a quantum phase transition, where the analytical techniques we have deployed become less reliable. Inelastic neutron scattering measurements would provide a direct experimental probe of fluctuations in BZRO and could be modeled using the methods presented here as well as more sophisticated approaches, if necessary. Incorporating finite-temperature dynamics using stochastic methods with colored noise would enable modeling of thermal fluctuations beyond mean-field theory and allow exploration of specific heat, dynamic correlations, and more quantitative finite-temperature magnetic response \cite{barker2019}. Additionally, studying the impact of structural disorder, ligand environment, and pressure may reveal tunable routes to control the balance between singlet formation and magnetic order in this and related quantum dimer systems. Finally, these 6H-perovskite dimer lattices may have interactions tuned to support novel skyrmion phases~\cite{Williams2025}.

\section*{Data Availability}
The inelastic neutron scattering data for Ba$_3$Zn(Ru$_{1-x}$Sb$_x$)$_2$O$_9$ that support the findings of this study are available at \href{https://doi.org/10.14461/oncat.data/2574871}{https://doi.org/10.14461/oncat.data/2574871}.

\begin{acknowledgments}
D.M.P. acknowledges Cristian Batista for valuable discussions and advice related to this work. A portion of this research used resources at the Spallation Neutron Source, a DOE Office of Science User Facility operated by Oak Ridge National Laboratory. Neutron beam time was allocated at CNCS and SEQUOIA under proposal number IPTS-29873. Work at Argonne National Laboratory (synthesis, neutron scattering) was sponsored by the U.S. Department of Energy, Office of Science, Basic Energy Sciences, Materials Science and Engineering Division.
\end{acknowledgments}

\appendix

\section{\label{sec:LT_appendix}Classical Luttinger--Tisza Analysis}

\begin{figure*}[hbt]
  \includegraphics[width=\linewidth]{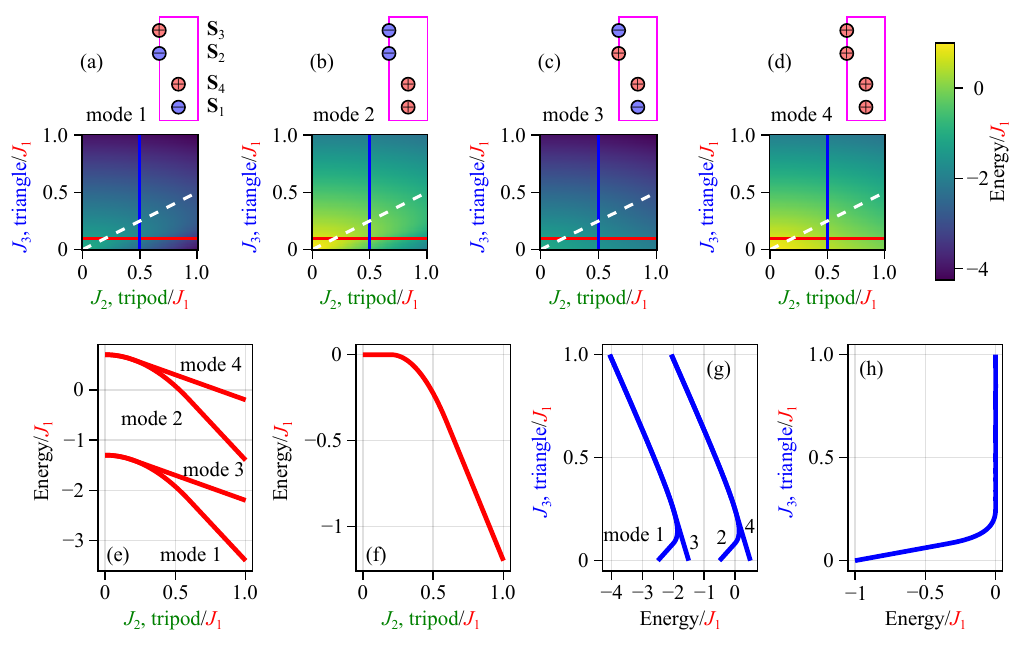}
  \caption{(Color online)
  Classical exchange energy landscape for the four LT modes on the 6H-perovskite dimer lattice, obtained by minimizing $\lambda_{\text{min}}(\mathbf{k})$ over momentum $\mathbf{k}$ for each point in exchange parameter space. (a--d): Minimal eigenvalue $\lambda_{\text{min}}(\mathbf{k}_0)$ of each mode across the $(J_2, J_3)$-plane with $J_4 = 0$. Dashed line marks $J_2/J_3 = 2$. The magenta boxes containing spin sites illustrate the relative signs of the spins in the eigenvector
mode, consistent with Figure~\ref{fig:LT-wavevecs}. Panels (e--h): One-dimensional cuts showing mode competition and degeneracy as a function of exchange ratios.}
  \label{fig:LT-energy-surfaces}
\end{figure*}

To identify candidate classical ground states and magnetic ordering
vectors, we apply the LT method~\cite{Luttinger1946,Litvin1974,Niemeijer1973} to the Heisenberg model on
the 6H-perovskite, AB-stacked dimer triangular lattice. The classical ground state is
determined by minimizing the lowest eigenvalue of the momentum-dependent
exchange matrix over the Brillouin zone. This analysis identifies the
ordering wavevector and associated eigenmode (4 possible here) as a
function of exchange parameters $J_2$ and
$J_3$, with $J_1$ setting the overall energy
scale and $J_4$=0. 

The position of each spin is labeled by a Bravais lattice vector $\mathbf{R}$ and a sublattice index $\alpha = 1,\ldots,4$, where the four sublattices correspond to the magnetic sites in the hexagonal unit cell (two vertically offset dimers).

The classical spin $\mathbf{S}_{\alpha}(\mathbf{R})$ is Fourier transformed as:
\begin{equation}
\mathbf{S}_{\alpha}(\mathbf{R}) = \frac{1}{\sqrt{N}} \sum_{\mathbf{k}} \mathbf{S}_{\alpha}(\mathbf{k}) e^{i \mathbf{k} \cdot \mathbf{R}}
\end{equation}
where $N$ is the number of unit cells, and $\mathbf{S}_{\alpha}(\mathbf{k})$ is the spin amplitude for sublattice $\alpha$ at wavevector $\mathbf{k}$. Substituting the Fourier-transformed spin expression into the Heisenberg Hamiltonian defined in Eq.~(1), we obtain the momentum-space form:
\begin{equation}
H = \sum_{\mathbf{k}} \sum_{\alpha,\beta} \mathbf{S}_{\alpha}^{\dagger}(\mathbf{k}) \cdot J_{\alpha \beta}(\mathbf{k}) \cdot \mathbf{S}_{\beta}(\mathbf{k})
\end{equation}
where $J_{\alpha \beta}(\mathbf{k})$ is a $4 \times 4$ Hermitian matrix in sublattice space that encodes the Fourier-transformed exchange interactions between sites $\alpha$ and $\beta$. The ground state is determined by minimizing the lowest eigenvalue $\lambda_{\text{min}}(\mathbf{k})$ of $J_{\alpha \beta}(\mathbf{k})$ over the Brillouin zone. For each wavevector $\mathbf{k}$, the eigenvalue equation
\begin{equation}
\mathbf{J}(\mathbf{k}) \cdot \mathbf{S}_{\mathbf{k}} = \lambda(\mathbf{k}) \cdot \mathbf{S}_{\mathbf{k}}
\end{equation}
yields four modes. The candidate ordering wavevector is the momentum $\mathbf{k}_{0}$ at which the lowest eigenvalue $\lambda_{\text{min}}(\mathbf{k})$ is minimized over the Brillouin zone. If the corresponding eigenmode satisfies the hard-spin constraint $|\mathbf{S}_{\alpha}(\mathbf{R})| = S$, the LT solution represents a valid classical ground state.

We now present the Fourier-transformed forms of the exchange interactions $J_1$, $J_2$, and $J_3$ as they enter the interaction matrix $J_{\alpha \beta}(\mathbf{k})$, neglecting $J_4$ in this analysis. The intradimer coupling $J_1$ connects spins within a dimer. The tripod exchange $J_2$ appears both within the unit cell and between neighboring cells and thus requires distinguishing between intra-cell and inter-cell terms. We denote the intra-cell version as $J_2'(\mathbf{k})$, which differs from $J_2(\mathbf{k})$ due to relative phase factors introduced by the Bravais lattice translations. The triangle interaction $J_3$ couples sites within the plane. The Fourier-transformed exchange contributions are:
\begin{equation}
\begin{aligned}
J_1(\mathbf{k}) &= J_1 \\
J_2(\mathbf{k}) &= J_2\left[ \cos 2\pi(h + l) + \cos(2\pi l) + \cos 2\pi(k - l) \right] \\
J_2'(\mathbf{k}) &= J_2\left[ \cos(2\pi h) + 1 + \cos(2\pi k) \right] \\
J_3(\mathbf{k}) &= 2J_3\left[ \cos 2\pi(h + k) + \cos(2\pi h) + \cos(2\pi k) \right]
\end{aligned}
\end{equation}
where $h$, $k$, and $l$ are Miller indices. Using the Fourier-transformed exchange interactions defined above, the interaction matrix $J_{\alpha \beta}(\mathbf{k})$ takes the form:
\begin{equation}
\mathbf{J}(\mathbf{k}) \begin{bmatrix} S_1 \\ S_2 \\ S_3 \\ S_4 \end{bmatrix} = 
\begin{bmatrix}
J_3(\mathbf{k}) & 0 & J_2(\mathbf{k}) & J_1(\mathbf{k}) \\
0 & J_3(\mathbf{k}) & J_1(\mathbf{k}) & J_2'(\mathbf{k}) \\
J_2(\mathbf{k}) & J_1(\mathbf{k}) & J_3(\mathbf{k}) & 0 \\
J_1(\mathbf{k}) & J_2'(\mathbf{k}) & 0 & J_3(\mathbf{k})
\end{bmatrix}
\begin{bmatrix} S_1 \\ S_2 \\ S_3 \\ S_4 \end{bmatrix}
\end{equation}

The $4 \times 4$ interaction matrix $J_{\alpha \beta}(\mathbf{k})$ admits a symbolic solution for its eigenvalues, which correspond to the four classical spin-wave modes at each wavevector $\mathbf{k}$. Solving the eigenvalue equation, we obtain:
\begin{equation}
\lambda_n(\mathbf{k}) = J_3(\mathbf{k}) + \frac{\sigma_1^{(n)}}{2}\left[ J_2(\mathbf{k}) + J_2'(\mathbf{k}) + \sigma_2^{(n)} \mathcal{T}(\mathbf{k}) \right]
\end{equation}
where the $\sigma_1$ and $\sigma_2$ terms keep track of different branches, with the square root discriminant:
\begin{equation}
\mathcal{T}(\mathbf{k}) = \sqrt{\left( J_2(\mathbf{k}) - J_2'(\mathbf{k}) \right)^2 + 4J_1^2}
\end{equation}
and we label $\lambda_n(\mathbf{k})$ as $n = 1,2,3,4$ corresponding to $(\sigma_1, \sigma_2) = (--), (-+), (+-), (++)$, respectively. The eigenvectors are:
\begin{equation}
\mathbf{S}^{(n)}(\mathbf{k}) = \begin{pmatrix}
\frac{\sigma_1^{(n)}\left[ J_2(\mathbf{k}) - J_2'(\mathbf{k}) \right] + \sigma_2^{(n)} \mathcal{T}(\mathbf{k})}{2J_1} \\
\sigma_1^{(n)} \\
\frac{J_2(\mathbf{k}) - J_2'(\mathbf{k}) + \sigma_1^{(n)}\sigma_2^{(n)} \mathcal{T}(\mathbf{k})}{2J_1} \\
1
\end{pmatrix}
\end{equation}

With these expressions, the LT phase diagram is computed by numerically minimizing $\lambda_{\text{min}}(\mathbf{k})$ while scanning $J_2/J_1$ and $J_3/J_1$ over $[0,1]$. For each $(J_2, J_3)$ pair, the Brillouin zone is searched for the wavevector $\mathbf{k}_0$ that minimizes $\lambda_{\text{min}}(\mathbf{k})$, and the corresponding eigenmode is identified. This yields a classical phase diagram in the $(J_2, J_3)$-plane delineating distinct ordering wavevectors and eigenmodes. For this region of parameter space:
\begin{equation}
\begin{aligned}
S^{(1)}(\mathbf{k}) &= [-, -, +, +] \\
S^{(2)}(\mathbf{k}) &= [+, -, -, +] \\
S^{(3)}(\mathbf{k}) &= [-, +, -, +] \\
S^{(4)}(\mathbf{k}) &= [+, +, +, +]
\end{aligned}
\end{equation}

\begin{figure}[th]
  \centering
  \includegraphics[width=\columnwidth]{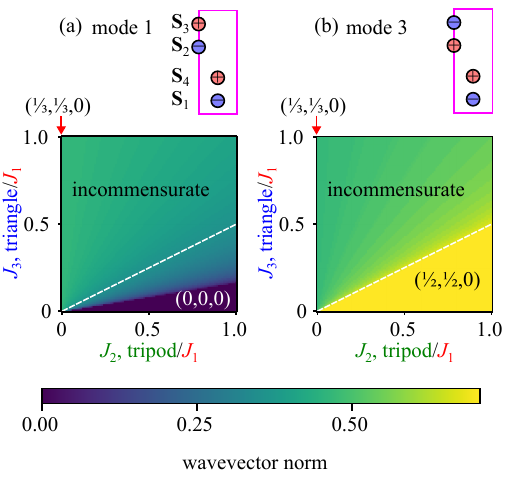}
  \caption{(Color online) Ground-state ordering wavevectors
\textbf{k}$_0$ obtained from LT analysis over
the ($J_2$,$J_3$) -plane with $J_4$=0 for the 6H-perovskite dimer
lattice. Color scale represents the norm of the minimized wavevector
\textbf{k}$_0$ = $(h,h,0)$, except along the diagonal line
$J_2/J_3$=2, where a degenerate manifold of
ordering wavevectors with the form ($h$, $\frac{1}{2} - h$,0) appears. Well-defined
commensurate phases are labeled, including $(0,0,0)$ for (a) mode
1, $(\frac{1}{2},\frac{1}{2},0)$ for (b) mode 3, and ($\frac{1}{3}$,$\frac{1}{3}$,0) in both when
$J_2=0$, while the remaining regions are incommensurate.
The magenta boxes and dot configurations illustrate the relative signs of the spins in the eigenvector
mode, consistent with Figure~\ref{fig:LT-energy-surfaces}.}
  \label{fig:LT-wavevecs}
\end{figure}

The lowest energy modes are labeled mode 1 and mode 3, are degenerate
for $J_2/J_3 < 2$, and mode 1 is the
ground state for $J_2/J_3 > 2$. In Fig.~\ref{fig:LT-wavevecs}, we map the ground-state wavevector $\mathbf{k}_0$ across the
$(J_2,J_3)$ parameter space. The LT ground
states are generally characterized by ordering vectors of the form
$(h,h,0)$, with $h$ evolving continuously across the phase diagram. However,
along the special line $J_2/J_3=2$, a
degenerate manifold of ordering vectors appears with the form $(h,
\frac{1}{2} - h,0)$, indicating enhanced frustration, although this degeneracy is
relieved when including quantum corrections as described in the
Appendix~\ref{sec:LSWT_appendix}. Commensurate phases are present in isolated regions: (0,0,0)
in mode 1, ($\frac{1}{2}$,$\frac{1}{2}$,0) in mode 3, and ($\frac{1}{3}$,$\frac{1}{3}$,0) in both when
$J_2=0$. Outside these boundaries, the ordering wavevector
becomes incommensurate.

\section{\label{sec:dimer_appendix}Dimer Mean-field Theory}

For each pair of exchange parameters $(J_{2}, J_{3})$, we use the ordering wavevector $\mathbf{k}_{0}$ from the LT analysis of mode 1 to evaluate the momentum-dependent exchange terms. The dimer Hamiltonian is diagonalized numerically, and the spin expectation values $\langle \mathbf{S}_{2} \rangle$ and $\langle \mathbf{S}_{3} \rangle$ are updated iteratively until self-consistency is achieved. This procedure yields the dimer mean-field ground state for each point in phase space.

From the converged solution, we extract the squared total spin, \( \langle \mathbf{S}_{\text{tot}}^{2} \rangle = \left\langle \left( \mathbf{S}_{2} + \mathbf{S}_{3} \right)^2 \right\rangle = S_{\text{tot}}(S_{\text{tot}} + 1) \), which reflects the degree of mixing among different multiplet states, including singlet, triplet, and higher-spin sectors, as well as the energy levels and the wavefunctions.

To identify the ordering temperature $T_\mathrm{N}$, we perform a binary search over temperature to locate the critical point where the self-consistent solution yields a nonzero local moment.
Specifically, we define the total staggered moment as the absolute value of the spin expectation on each site in the dimer and take the onset of a finite value $\lvert \langle S_z \rangle \rvert > 10^{-4}$ as the indicator of spontaneous symmetry breaking.

\section{\label{sec:LSWT_appendix}Linear Spin-wave Theory}

In the classical limit, the exchange model with $J_{2}/J_{3} = 2$ exhibits a continuous degeneracy of ground states along the line $\mathbf{k}_{0} = (h, \tfrac{1}{2} - h, 0)$, reflecting the geometric frustration of the lattice. The magnon free energy, $F = E_{\text{ZPE}} - TS$, is dominated by the zero-point energy at low temperatures. 

\begin{figure}[!h]
    \centering
    \includegraphics[width=3.4in]{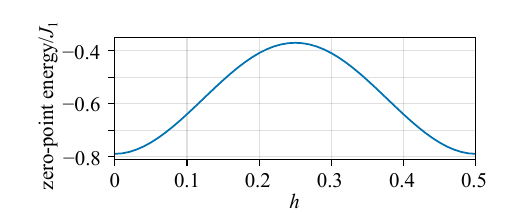}
    \caption{(Color online)
    \label{fig:ZPEvsh}
    Traditional LSWT zero-point energy versus $h$ for $\mathbf{k}_{0} = (h, \tfrac{1}{2} - h, 0)$ on the 6H-perovskite dimer lattice with $J_{2}/J_1 = 0.5$, $J_{3}/J_1 = 0.25$, and $J_{4} = 0$.}
\end{figure}

Using LSWT with a single-$\mathbf{k}$ spiral, we compute the zero-point energy for $J_2/J_1 = 0.5$, $J_{3}/J_1 = 0.25$, and $J_{4} = 0$, as shown in Figure~\ref{fig:ZPEvsh}. The resulting energy landscape reveals that quantum fluctuations lift the classical degeneracy, favoring a unique ordering vector with $\mathbf{k}_{0} = (\tfrac{1}{2}, 0, 0)$ or $(0, \tfrac{1}{2}, 0)$, illustrating a clear case of quantum order-by-disorder in the 6H-perovskite dimer lattice.

\section{\label{sec:DFT_appendix}Density Functional Theory}

The DFT calculations were performed using the Vienna \emph{Ab initio} Simulation Package (VASP)~\cite{Kresse1996_CMS,Kresse1996_PRB}, employing the projector augmented wave (PAW) method~\cite{Bloechl1994,Kresse1999} and the Perdew--Burke--Ernzerhof (PBE) generalized gradient approximation (GGA) functional~\cite{Perdew1996}. The plane-wave kinetic energy cutoff was set to 520\,eV, and all calculations used the ``Accurate'' precision setting with an electronic convergence criterion of $1 \times 10^{-8}$\,eV.
Valence electrons were described using the standard \texttt{PAW\_PBE} pseudopotentials distributed with VASP. Specifically, Ru was modeled with the \texttt{Ru\_pv} potential (4p$^6$5s$^1$4d$^7$, 14 valence electrons), Sb with \texttt{Sb} (5s$^2$5p$^3$, 5 electrons), Ba with \texttt{Ba\_sv} (5s$^2$5p$^6$6s$^2$, 10 electrons), Zn with \texttt{Zn} (3d$^{10}$4s$^2$, 12 electrons), and O with \texttt{O} (2s$^2$2p$^4$, 6 electrons). Ca was modeled with \texttt{Ca\_sv} (3s$^2$3p$^6$4s$^2$, 10 valence electrons) where applicable.
To account for strong electron correlations in the localized Ru 4d orbitals, the GGA+$U$ method was applied using the rotationally invariant Dudarev approach with $U_\mathrm{eff} = U - J$. This correction mitigates self-interaction error and enhances electron localization in the Ru$^{5+}$ ($S = \tfrac{3}{2}$) ions, preventing spurious metallic behavior.

The first set of calculations was performed using a single crystallographic unit cell of BZRO with a $\Gamma$-centered $8 \times 8 \times 3$ \emph{k}-point mesh. A Hubbard correction of $U_\mathrm{eff} = U - J = 2.3$\,eV was applied to the Ru 4d orbitals, based on previously successful values of $U = 3.00$\,eV and $J = 0.7$\,eV used in DFT studies of Ba$_3$CoRu$_2$O$_9$~\cite{Streltsov2013}.
The initial atomic coordinates were taken from a published crystallographic information file (CIF)~\cite{Beran2015}. These calculations used collinear spin configurations, with initial moments set to $\pm 3\,\mu_\mathrm{B}$ on each Ru atom. Four different magnetic arrangements were considered, distinguished by the spin orientations of the four Ru atoms at fractional $z$-coordinates (0.1551, 0.8449, 0.6551, 0.3449). This ordering is used in the text to label the magnetic structures. Structural relaxations were performed until all atomic forces were less than 0.01\,eV/\AA. An analogous procedure was undertaken for BCRO.

The four shortest Ru--Ru superexchange pathways are summarized in Table~\ref{tab:superexchange}, indexed according to the labels shown in Fig.~1. Distances are provided both from the initial crystallographic CIF and from DFT-relaxed structures in the lowest-energy magnetic configuration for each compound. Relaxation leads to only minor adjustments in lattice parameters for BZRO compared to the experimental CIF, but comparison between BZRO and BCRO reveals key structural changes induced by B-site substitution. The nonmagnetic B-site cation occupies the space between the triangular dimer planes, effectively acting as a spacer. The larger ionic radius of Ca$^{2+}$ (100\,pm) compared to Zn$^{2+}$ (74\,pm) results in greater separation between adjacent dimer layers~\cite{Shannon1976}. This structural change both pulls dimers apart along the $c$-axis and in the $ab$-plane and slightly compresses them within the Ru$_2$O$_9$ unit, though the former effect dominates.

These subtle geometric changes directly impact the superexchange pathways when moving from BCRO to BZRO: $J_1$ is expected to weaken due to increased intradimer Ru--Ru distance, while $J_2$--$J_4$ are anticipated to strengthen owing to the closer proximity of neighboring dimers. Altogether, the substitution from Ca to Zn tunes the system from more weakly coupled dimers toward a more strongly interacting regime.

\begin{table}[b]
\caption{\label{tab:superexchange}%
Ru--Ru superexchange distances (in \AA) for BCRO and BZRO. Relaxed DFT values correspond to the lowest-energy magnetic structure ($B_1$). Coordination number $z$ indicates the number of equivalent bonds per Ru site.
}
\begin{ruledtabular}
\begin{tabular}{lcccc}
Exchange & BZRO (CIF) & BZRO (relaxed) & BCRO (relaxed) & $z$ \\
\hline
$J_1$ & 2.682 & 2.701 & 2.641 & 1 \\
$J_2$ & 5.501 & 5.493 & 5.733 & 3 \\
$J_3$ & 5.755 & 5.760 & 5.881 & 6 \\
$J_4$ & 6.349 & 6.362 & 6.447 & 6 \\
\end{tabular}
\end{ruledtabular}
\end{table}

The \texttt{Sunny.jl} analysis suite was used to calculate the symmetry-allowed exchange matrices for the four shortest Ru--Ru bonds. These matrices capture the most general bilinear interactions between spins on each exchange path, constrained by the local symmetry of the bond. The resulting forms are:

\begin{equation}
\begin{aligned}
\overleftrightarrow{J}_{1} &= 
\begin{bmatrix}
A & 0 & 0 \\
0 & A & 0 \\
0 & 0 & B
\end{bmatrix}, \quad
\overleftrightarrow{J}_{2} = 
\begin{bmatrix}
A & 0 & 0 \\
0 & B & D \\
0 & D & C
\end{bmatrix}, \\
\overleftrightarrow{J}_{3} &= 
\begin{bmatrix}
A & F & -E \\
-F & B & D \\
E & D & C
\end{bmatrix}, \quad
\overleftrightarrow{J}_{4} = 
\begin{bmatrix}
A & F & D \\
-F & B & E \\
D & -E & C
\end{bmatrix}.
\end{aligned}
\end{equation}
The parameters $A$, $B$, $C$, $D$, $E$, and $F$ are independent for each matrix and represent the allowed components of the exchange tensor under symmetry. Notably, $\overleftrightarrow{J}_3$ and $\overleftrightarrow{J}_4$ include symmetry-allowed antisymmetric (Dzyaloshinskii–Moriya) exchange terms. In the analysis that follows, we restrict ourselves to the isotropic Heisenberg limit and consider only the scalar exchange contributions, although the possibility of anisotropic exchanges suggests the potential for tuning these systems to support more rich physics. 

The four calculated spin configurations of the single-cell DFT are labeled using ``$+$'' and ``$-$'' signs, following the order of Ru atoms listed earlier. The ``$++++$'' ($A_1$) configuration corresponds to all magnetic moments aligned ferromagnetically. The ``$+-+-$'' ($B_1$) configuration exhibits antiferromagnetic alignment both within each Ru$_2$O$_9$ structural dimer and between adjacent dimer layers. The ``$+--+$'' ($C_1$) configuration features parallel spins within the dimers but antiparallel alignment across dimer layers. Conversely, the ``$++--$'' ($D_1$) structure displays antiferromagnetic alignment within each dimer while maintaining ferromagnetic alignment between layers.

These configurations enable mapping onto a classical Heisenberg spin Hamiltonian with four superexchange couplings, $J_1$ through $J_4$, corresponding to the dominant Ru--Ru exchange paths identified earlier. The total classical superexchange energy for each configuration is summarized in Table~\ref{tab:superexchange_energy}. All four structures have the same $J_3$ contribution.

\begin{table}[h]
\caption{\label{tab:superexchange_energy}Spin configurations and corresponding classical Heisenberg superexchange energies for four collinear magnetic structures.}
\begin{ruledtabular}
\begin{tabular}{lc>{\scriptsize}c}
Label & Magnetic Structure & Superexchange Energy \\
\hline
$A_1$ & $++++$ & $+12J_4S^2 + 12J_3S^2 + 6J_2S^2 + 2J_1S^2$ \\
$B_1$ & $+{-}+{-}$ & $-12J_4S^2 + 12J_3S^2 - 6J_2S^2 - 2J_1S^2$ \\
$C_1$ & $+--+$ & $+12J_4S^2 + 12J_3S^2 - 6J_2S^2 + 2J_1S^2$ \\
$D_1$ & $++--$ & $-12J_4S^2 + 12J_3S^2 + 6J_2S^2 - 2J_1S^2$ \\
\end{tabular}
\end{ruledtabular}
\end{table}

Table~\ref{tab:dft_summary} summarizes the DFT total energy (relative to $B_1$), band gap $\Delta_0$, and local Ru magnetic moment $m_\mathrm{Ru}$ for each configuration in both BZRO and BCRO. In all cases, the $B_1$ configuration is the lowest in energy. The calculated Ru magnetic moment is significantly reduced from the ideal high-spin value of $3\,\mu_\mathrm{B}$ to $\approx 2\,\mu_\mathrm{B}$ due to strong hybridization with oxygen and the extended 4$d$ orbital character. No Hubbard $U_\mathrm{eff}$ was applied to the oxygen sites, though this could improve accuracy at the cost of introducing more parameters.

\begin{table}[h]
\caption{\label{tab:dft_summary}Total DFT energy (relative to $B_1$), band gap $\Delta_0$, and local Ru magnetic moment $m_\mathrm{Ru}$ for four collinear spin configurations in single-unit-cell BZRO and BCRO.}
\begin{ruledtabular}
\begin{tabular}{ccccc}
Structure & Label & $E$ (meV) & $\Delta_0$ (eV) & $m_\mathrm{Ru}$ ($\mu_\mathrm{B}$) \\
\hline
\multicolumn{5}{c}{BZRO} \\
$++++$ & $A_1$ & 369.24 & 0.4 & 1.98 \\
$+-+-$ & $B_1$ & 0.00 & 1.1 & 1.87 \\
$+--+$ & $C_1$ & 174.86 & 0.9 & 1.90 \\
$++--$ & $D_1$ & 168.53 & 0.8 & 1.93 \\
\multicolumn{5}{c}{BCRO} \\
$++++$ & $A_1$ & 327.06 & 0.4 & 2.01 \\
$+-+-$ & $B_1$ & 0.00 & 1.2 & 1.89 \\
$+--+$ & $C_1$ & 157.44 & 0.7 & 1.98 \\
$++--$ & $D_1$ & 103.21 & 0.9 & 1.91 \\
\end{tabular}
\end{ruledtabular}
\end{table}

Exchange constants were estimated from total energy differences between the single-cell collinear magnetic configurations using the expressions in Table~\ref{tab:superexchange_energy}, assuming a spin magnitude of $S = \tfrac{3}{2}$. Since these configurations preserve in-plane symmetry, they are insensitive to $J_3$, and due to linear dependence among the equations, only the combination $J_1 + 6J_4$ can be extracted. Values obtained using the ferromagnetic ($A_1$) and antiferromagnetic ($C_1$) configurations differ by approximately 10\%, likely due to greater deviation from the true ground state electronic structure in the ferromagnetic case.

The extracted values are summarized in Table~\ref{tab:exchange_extracted}, which shows a modest reduction in $J_1 + 6J_4$ and a more significant suppression of $J_2$ upon Zn-to-Ca substitution in the B-site.

\begin{table}[h]
\caption{\label{tab:exchange_extracted}Exchange constants (in meV) extracted from single-cell total energy differences (in parentheses where delineated) between spin configurations for BZRO and BCRO, assuming $S = \tfrac{3}{2}$.}
\begin{ruledtabular}
\begin{tabular}{lccc}
Compound & $J_1 + 6J_4$ ($C_1$) & $J_1 + 6J_4$ ($A_1$) & $J_2$ \\
\hline
BZRO & 19.43 & 22.31 & 6.24 \\
BCRO & 17.49 & 24.88 & 3.82 \\
\end{tabular}
\end{ruledtabular}
\end{table}

For the BZRO compound in its $B_1$ atomic and magnetic ground-state configuration, spin--orbit coupling (\texttt{LSORBIT = .TRUE.}) and non-collinear magnetism (\texttt{LORBMOM = .TRUE.}) were included in the DFT calculations to evaluate the relative energies of spin orientations along the $x$-, $y$-, and $z$-axes. The $x$ and $y$ directions were found to be degenerate in energy, while the $z$-axis was energetically favored, indicating easy-axis anisotropy. The corresponding single-ion uniaxial anisotropy constant ($DS_z^2$) was estimated to be $D = -0.03$\,meV.

The second set of DFT calculations was performed for crystallographic cells of BCRO and BZRO doubled along the crystallographic $a$-axis. A $\Gamma$-centered mesh with $(4 \times 8 \times 3)$ $k$-points was used. In this series, we scanned $U_{\mathrm{eff}}$ for the Ru 4$d$ electrons. The atomic positions were fixed to those of the $B_1$ relaxed structure described earlier. Eight collinear magnetic structures were considered, each initialized with $\pm3$\,$\mu_\mathrm{B}$ per Ru ion. The fractional $z$-coordinates of the eight Ru atoms were $(\overline{0.155}, 0.155, \overline{0.845}, 0.845, \overline{0.654}, 0.654, \overline{0.345}, 0.345)$; overlined entries indicate atoms in the original unit cell, while the others belong to the doubled portion. This ordering is used to label spin configurations.

The computed magnetic structures and their superexchange energies, based on a generic Heisenberg Hamiltonian with exchanges $J_1$ through $J_4$, are summarized in Table~\ref{tab:doublecell_superexchange}. Structures $A_2$--$D_2$ are doubled versions of $A_1$--$D_1$, while $A_2$K--$D_2$K introduce sign flips in the duplicated cell, allowing sensitivity to $J_3$.

\begin{table}[h]
\caption{\label{tab:doublecell_superexchange}Spin configurations and corresponding classical Heisenberg superexchange energies for eight collinear magnetic structures calculated in the doubled-unit-cell DFT. Overlined symbols indicate spins on atoms in the original unit cell.}
\begin{ruledtabular}
\begin{tabular}{ll>{\scriptsize}c}
Label & Magnetic Structure & Superexchange Energy \\
\hline
$A_2$ & $\overline{+}+\,\overline{+}+\,\overline{+}+\,\overline{+}+$ &
$+24J_4S^2 + 24J_3S^2 + 12J_2S^2 + 4J_1S^2$ \\
$B_2$ & $\overline{+}+\,\overline{-}-\,\overline{+}+\,\overline{-}-$ &
$-24J_4S^2 + 24J_3S^2 - 12J_2S^2 - 4J_1S^2$ \\
$C_2$ & $\overline{+}+\,\overline{-}-\,\overline{-}-\,\overline{+}+$ &
$+24J_4S^2 + 24J_3S^2 - 12J_2S^2 + 4J_1S^2$ \\
$D_2$ & $\overline{+}+\,\overline{+}+\,\overline{-}-\,\overline{-}-$ &
$-24J_4S^2 + 24J_3S^2 + 12J_2S^2 - 4J_1S^2$ \\
$A_2$K & $\overline{+}-\,\overline{+}-\,\overline{+}-\,\overline{+}-$ &
$-8J_4S^2 - 8J_3S^2 + 4J_2S^2 + 4J_1S^2$ \\
$B_2$K & $\overline{+}-\,\overline{-}+\,\overline{+}-\,\overline{-}+$ &
$+8J_4S^2 - 8J_3S^2 - 4J_2S^2 - 4J_1S^2$ \\
$C_2$K & $\overline{+}-\,\overline{-}+\,\overline{-}+\,\overline{+}-$ &
$-8J_4S^2 - 8J_3S^2 - 4J_2S^2 + 4J_1S^2$ \\
$D_2$K & $\overline{+}-\,\overline{+}-\,\overline{-}+\,\overline{-}+$ &
$+8J_4S^2 - 8J_3S^2 + 4J_2S^2 - 4J_1S^2$ \\
\end{tabular}
\end{ruledtabular}
\end{table}

Numerical fits to the Heisenberg model were performed using the configuration energies of Table~\ref{tab:doublecell_superexchange} and DFT-computed energies for each configuration. The fitted energies compared to the DFT energies are listed in Table~\ref{tab:fit_quality} for both BZRO and BCRO at $U_{\mathrm{eff}} = 2.3$\,eV, with the largest deviation between the VASP energy and the Heisenberg model energy being less than 10\,meV for the full double-cell. This style of fit was repeated for each $U_{\mathrm{eff}}$ value to extract the exchange constants reported in Tables~\ref{tab:BZRO_exchanges} and \ref{tab:BCRO_fit_exchanges}. The extracted offset energy $E_0$ correlates with the $U_{\mathrm{eff}}$ used in the DFT. Comparison with experimental results from BCRO, modeled assuming isolated dimers, provides a useful benchmark: historical INS yields $J_1 \approx 26$\,meV, our re-analysis gives 30.6\,meV, and magnetic susceptibility fits yield $J_1 \approx 29$\,meV. These comparisons suggest that $U_{\mathrm{eff}}$ values between 2.3\,eV and 2.8\,eV best capture the magnetic behavior of BZRO and BCRO. In this range, $J_1$ for BZRO is found to be significantly weaker than $J_1$ for BCRO, consistent with the greater separation within the Ru$_2$O$_9$ structural dimers. Conversely, the interdimer couplings $J_2$--$J_4$ show systematic enhancement in BZRO compared to BCRO, consistent with the shorter Ru--Ru interdimer distances.

\begin{table}[h]
\caption{\label{tab:fit_quality}Comparison of DFT energies ($E_{\mathrm{VASP}}$) to Heisenberg model fit energies ($E_{\mathrm{model}}$) for the double-cell calculations at $U_{\mathrm{eff}} = 2.3$\,eV, in units of meV, and relative to the lowest energy state.}
\begin{ruledtabular}
\begin{tabular}{llrrr}
Compound & Label & $E_{\mathrm{VASP}}$ & $E_{\mathrm{model}}$ & Difference \\
\hline
BZRO & $A_2$   & 754.38 & 746.72 & 7.66 \\
     & $B_2$   &   0.00 &  $-5.05$ & 5.05 \\
     & $C_2$   & 364.97 & 372.63 & $-7.66$ \\
     & $D_2$   & 363.99 & 369.04 & $-5.05$ \\
     & $A_2$K  & 514.20 & 516.63 & $-2.42$ \\
     & $B_2$K  &  59.18 &  53.78 & 5.40 \\
     & $C_2$K  & 394.34 & 391.92 & 2.42 \\
     & $D_2$K  & 173.07 & 178.47 & $-5.40$ \\
BCRO & $A_2$   & 899.47 & 889.66 & 9.81 \\
     & $B_2$   &   0.00 & $-7.41$ & 7.41 \\
     & $C_2$   & 563.70 & 573.51 & $-9.81$ \\
     & $D_2$   & 301.32 & 308.74 & $-7.42$ \\
     & $A_2$K  & 652.48 & 653.97 & $-1.49$ \\
     & $B_2$K  &  11.01 &   5.31 & 5.70 \\
     & $C_2$K  & 550.08 & 548.59 & 1.49 \\
     & $D_2$K  & 104.99 & 110.69 & $-5.70$ \\
\end{tabular}
\end{ruledtabular}
\end{table}

\begin{table}[h]
\caption{\label{tab:BZRO_exchanges}Exchange constants (in meV) extracted from BZRO double-cell DFT calculations as a function of $U_{\mathrm{eff}}$.}
\begin{ruledtabular}
\begin{tabular}{rrrrrr}
$U_{\mathrm{eff}}$ (eV) & $E_0$ & $J_1$ & $J_2$ & $J_3$ & $J_4$ \\
\hline
0.0 & $-415496$ & 45.02 & 11.12 & 1.88 & 0.60 \\
1.8 & $-401693$ & 23.16 & 7.60 & 1.30 & 0.31 \\
2.3 & $-397955$ & 19.33 & 6.93 & 1.19 & 0.27 \\
2.8 & $-394253$ & 16.10 & 6.34 & 1.09 & 0.24 \\
3.3 & $-390585$ & 13.33 & 5.83 & 1.01 & 0.22 \\
3.8 & $-386953$ & 10.96 & 5.36 & 0.93 & 0.19 \\
\end{tabular}
\end{ruledtabular}
\end{table}

\begin{table}[h]
\caption{\label{tab:BCRO_fit_exchanges}Exchange constants (in meV) extracted from BCRO double-cell DFT calculations as a function of $U_{\mathrm{eff}}$.}
\begin{ruledtabular}
\begin{tabular}{rrrrrr}
$U_{\mathrm{eff}}$ (eV) & $E_0$ & $J_1$ & $J_2$ & $J_3$ & $J_4$ \\
\hline
0.0 & $-432073$ & 70.35 & 8.53 & 1.18 & $-0.66$ \\
1.8 & $-418285$ & 36.26 & 6.44 & 1.72 & 0.31 \\
2.3 & $-414558$ & 30.71 & 5.85 & 1.55 & 0.26 \\
2.8 & $-410872$ & 26.02 & 5.35 & 1.40 & 0.22 \\
3.3 & $-407224$ & 22.06 & 4.89 & 1.28 & 0.20 \\
3.8 & $-403615$ & 18.68 & 4.49 & 1.17 & 0.17 \\
\end{tabular}
\end{ruledtabular}
\end{table}

To further isolate the contributions of individual exchange interactions and set the stage for an additional experimental study in the next section, a third set of DFT calculations was performed on magnetically dilute BZRSO with low Ru concentration. A $3\times2\times1$ supercell was constructed using the DFT-relaxed BZRO ground-state structure. All but one or two Ru atoms were replaced by Sb to yield isolated monomers or dimers, respectively. Dimers were constructed to probe each of the four superexchange pathways. Both ferromagnetic and antiferromagnetic spin alignments were considered for each dimer configuration, enabling direct evaluation of the exchange energy via subtraction, without the need to fit a system of equations.

All calculations employed spin-polarized DFT (\texttt{ISPIN = 2}) using the projector-augmented wave (PAW) method implemented in VASP. Magnetic moments were initialized on Ru sites (e.g., \texttt{MAGMOM = 0 0 3.0 357*0}), and convergence was achieved with high precision (\texttt{EDIFF = 1E-8}) and an energy cutoff of 520\,eV. Spin--orbit coupling (\texttt{LSORBIT = .TRUE.}) and non-collinear magnetism (\texttt{LORBMOM = .TRUE.}) were included. On-site Coulomb interactions were treated using Dudarev's formalism (\texttt{LDAUTYPE = 2}) with $U = 3.5$\,eV and $J = 0.7$\,eV for Ru. The tetrahedron method with Blöchl corrections (\texttt{ISMEAR = -5}) was used, symmetry was disabled (\texttt{ISYM = -1}), and a $\Gamma$-centered $3\times3\times3$ $k$-point mesh was applied.

The resulting exchange constants, extracted from total energy differences between ferromagnetic and antiferromagnetic configurations, are presented in Table~\ref{tab:BZRSO_exchanges} for several values of $U_{\mathrm{eff}}$. These values closely agree with those obtained for the parent BZRO system under similar conditions, although $J_2$--$J_4$ are somewhat enhanced in the isolated dimer BZRSO calculations. Additionally, a monomer calculation yielded a single-ion anisotropy constant of $D = -0.16$\,meV, confirming an easy-axis preference along the crystallographic $c$-axis, and indicating stronger anisotropy for isolated Ru monomers compared to the BZRO parent compound.

\begin{table}[h]
\caption{\label{tab:BZRSO_exchanges}Exchange constants (in meV) extracted from magnetically dilute BZRSO DFT calculations as a function of $U_{\mathrm{eff}}$.}
\begin{ruledtabular}
\begin{tabular}{rrrrr}
$U_{\mathrm{eff}}$ (eV) & $J_1$ & $J_2$ & $J_3$ & $J_4$ \\
\hline
0.0 & 45.08 & 12.03 & 2.28 & 0.52 \\
1.8 & 22.92 & 8.11 & 1.49 & 0.33 \\
2.3 & 19.10 & 7.39 & 1.35 & 0.30 \\
2.8 & 15.88 & 6.77 & 1.23 & 0.27 \\
3.3 & 13.13 & 6.21 & 1.13 & 0.24 \\
3.8 & 10.78 & 5.71 & 1.04 & 0.22 \\
\end{tabular}
\end{ruledtabular}
\end{table}

\section{\label{sec:dilute_appendix}Neutron Spectroscopy of Ba$_3$Zn(Ru$_{1-x}$Sb$_x$)$_2$O$_9$}

Neutron spectroscopy measurements were performed at the Spallation Neutron Source (SNS) using two complementary instruments. Low-energy excitations were probed using the Cold Neutron Chopper Spectrometer (CNCS), while higher-energy spin excitations were accessed via the SEQUOIA Fine-Resolution Fermi Chopper Spectrometer. For CNCS, the double-disk chopper was operated at 300\,Hz with the high-flux opening. For SEQUOIA, the high-flux Fermi chopper setting was used with a speed of 240\,Hz. Multiple incident energies ($E_i$) were employed to resolve features in the excitation spectrum: 1.00 meV, 3.32 meV, and 12 meV on CNCS, and 100 meV on SEQUOIA. Data were collected at $T = T_{base}$ and $T = 300$ K, where $T_{base} = 2$~K for CNCS and $T_{base} = 5$~K for SEQUOIA. A difference spectrum ($T_{base}$ minus 300 K) was used to suppress temperature-independent background and isolate magnetic scattering.

Fits to the neutron spectra were performed using ED of finite magnetic clusters including exchange interactions up to $J_4$, as defined in Table~\ref{tab:cluster_stats}. The dynamic structure factor for a cluster at temperature $T$ is given by~\cite{Furrer1979}:

\begin{eqnarray}
S(|\mathbf{Q}|, \hbar\omega) &=
\sum_{i,f} p_i
\sum_{j,j'} \sum_{\alpha = x,y,z}
\left\langle i \middle| S_j^\alpha \middle| f \right\rangle
\left\langle f \middle| S_{j'}^\alpha \middle| i \right\rangle \notag \\
&\quad \times F_j(|\mathbf{Q}|) F_{j'}(|\mathbf{Q}|)
\, \mathrm{sinc}(Q R_{jj'})
\label{Sqw}
\end{eqnarray}
where $\lvert i \rangle$ and $\lvert f \rangle$ are the initial and final eigenstates of the cluster transition,
$p_i = \frac{e^{-(E_i - E_0)/k_B T}}{Z}$ is the Boltzmann population of state $i$ with ground-state energy $E_0$ and partition function $Z$,
$S_j^{\alpha}$ is the spin operator component ($\alpha \in {x, y, z}$) on site $j$,
$F_j(|\mathbf{Q}|)$ is the magnetic form factor of ion $j$,
$R_{jj'}$ is the distance between sites $j$ and $j'$, and
$\mathrm{sinc}(|\mathbf{Q}| R_{jj'}) = \frac{\sin(|\mathbf{Q}| R_{jj'})}{|\mathbf{Q}| R_{jj'}}$
is the interference term arising from powder averaging over the cluster geometry.

\begin{figure*}[htb]
\includegraphics[width=\linewidth]{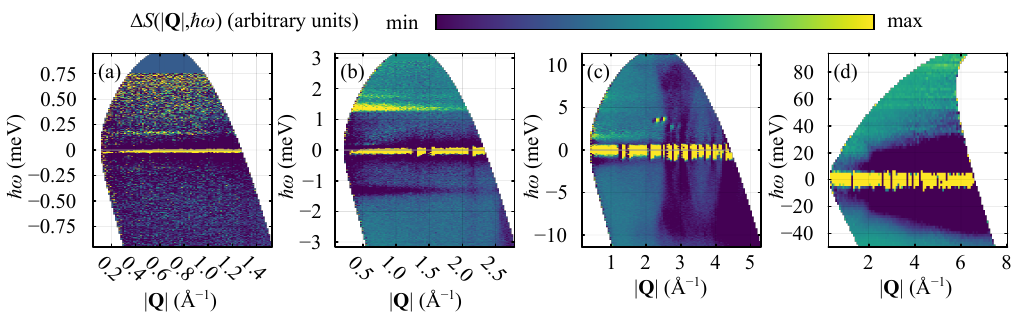}
\caption{(Color online)
\label{fig:INS-2d}Neutron spectroscopy results for BZRSO showing $\Delta S(|\mathbf{Q}|,\hbar\omega) = S(|\mathbf{Q}|,\hbar\omega, T_{base}) - S(|\mathbf{Q}|,\hbar\omega, 300\,\mathrm{K})$ as intensity heatmaps in $|\mathbf{Q}|$--$\hbar\omega$ space. (a) $E_i = 1.00$\,meV (CNCS), color limits: ($-0.2$, $0.5$); (b) $E_i = 3.32$\,meV (CNCS), color limits: ($-2.5$, $2.5$); (c) $E_i = 12.0$\,meV (CNCS), color limits: ($-2.5$, $2.5$); (d) $E_i = 100$\,meV (SEQUOIA), color limits: ($-0.05$, $0.03$).}
\end{figure*}

To interpret the INS data in terms of cluster excitations, we employ an exact diagonalization (ED)-based approach to compute the powder-averaged dynamic structure factor $S(|\mathbf{Q}|,\hbar\omega)$ for each cluster geometry. For a given spin cluster, the full Hamiltonian is numerically diagonalized to obtain eigenvalues and eigenstates. Thermal population of the initial states is computed using Boltzmann statistics, with a global shift such that the ground-state energy is zero to avoid numerical instability. The INS cross-section is calculated by summing over all thermally accessible transitions, each weighted by the corresponding spin matrix elements. These transitions are convolved with a Gaussian energy resolution profile (based on PyChop models within \textsc{Mantid}~\cite{Arnold2014}) and a Lorentzian broadening to account for finite lifetime effects. Momentum dependence enters via both the magnetic form factor and an interference term that reflects the spin cluster geometry, using spherical Bessel function approximations for powder averaging. Contributions from different clusters are summed incoherently.

To estimate the statistical distribution of magnetic clusters in Ba$_3$Zn(Ru$_{1-x}$Sb$_x$)$_2$O$_9$, we developed a numerical simulation based on random sampling of Ru$^{5+}$ occupancy within a crystallographic supercell. Starting from the relaxed BZRO structure, large supercells (e.g., $50\times50\times50$ unit cells) were generated, and Ru$^{5+}$ ions were probabilistically assigned to B-sites according to the nominal stoichiometry. A depth-first search algorithm was applied to identify connected clusters of Ru$^{5+}$ ions, where connectivity was defined by a distance cutoff equal to the longest bond length in the $J_4$ interaction. Each cluster was categorized as a monomer, dimer, trimer, or larger structure, with dimers and trimers further classified by geometry and pairwise Ru$^{5+}$ distances. The results are summarized in Table~\ref{tab:cluster_stats}, including the dominant exchange paths for each cluster type. This analysis provided quantitative estimates of relative cluster populations, guiding interpretation of neutron data and identifying which excitations could be attributed to specific clusters.

\begin{table}[h]
\caption{\label{tab:cluster_stats}Statistical cluster distribution in Ba$_3$Zn(Ru$_{1-x}$Sb$_x$)$_2$O$_9$ based on supercell simulations. The $x = 0.1$ and $x = 0.05$ results used $50\times50\times50$ supercells (500,000 B-sites), while the $x = 0.025$ result used an $80\times80\times80$ supercell (2,048,000 B-sites).}
\begin{ruledtabular}
\begin{tabular}{lcccc}
Cluster Type &
\shortstack{\\\% \\ $x=0.1$} &
\shortstack{\\\% \\ $x=0.05$} &
\shortstack{\\\% \\ $x=0.025$} &
Exchanges \\
\hline
monomer        & 52.1 & 69.0 & 82.5 & -- \\
dimer ($J_1$)  & 2.04 & 1.63 & 0.87 & $J_1$ \\
dimer ($J_2$)  & 2.86 & 3.23 & 2.28 & $J_2$ \\
dimer ($J_3$)  & 7.78 & 7.05 & 5.16 & $J_3$ \\
dimer ($J_4$)  & 6.55 & 6.65 & 5.20 & $J_4$ \\
trimer         & 1.06 & 0.75 & 0.32 & $2\times J_3$ \\
trimer         & 0.32 & 0.17 & 0.07 & $3\times J_3$ \\
trimer         & 1.09 & 0.68 & 0.23 & $J_1 + J_3$ \\
trimer         & 1.13 & 0.79 & 0.34 & $J_2 + J_3$ \\
trimer         & 0.38 & 0.24 & 0.30 & $J_3 + 2\times J_2$ \\
trimer         & 0.43 & 0.28 & 0.27 & $J_2 + J_1$ \\
trimer (misc.) & 5.04 & 3.84 & 1.70 & $J_4$ and others \\
tetramer       & 5.47 & 2.90 & 0.76 & Unclassified \\
pentamer       & 3.68 & 1.38 & 0.19 & Unclassified \\
$n$-mer ($n>5$) & 10.08 & 1.42 & 0.08 & Unclassified \\
\end{tabular}
\end{ruledtabular}
\end{table}

Polycrystalline BZRSO with a target of $5\,\%$ Ru on the B-site was synthesized by conventional solid-state reaction from BaCO$_3$, ZnO, RuO$_2$, and Sb$_2$O$_5$. The powders were mixed and calcined for $24$~h at $800~^\circ\mathrm{C}$. The mixture was then fired twice for $10$~h at $1000~^\circ\mathrm{C}$, followed by a final firing at $1200~^\circ\mathrm{C}$ for $4$~h, with regrinding between each step.

Representative intensity maps as a function of energy transfer ($\hbar\omega$) and momentum transfer magnitude ($|\mathbf{Q}|$) are shown in Fig.~\ref{fig:INS-2d}. For $E_i = 1.00$\,meV [Fig.~\ref{fig:INS-2d}(a)], a flat mode near 0.2\,meV is observed. For $E_i = 3.32$\,meV [Fig.~\ref{fig:INS-2d}(b)], modes appear near 1.3\,meV and 1.7\,meV, with thermal population of excited states evident as negative intensity on the energy gain side. For $E_i = 12$\,meV [Fig.~\ref{fig:INS-2d}(c)], the difference spectrum $\Delta S(Q,\hbar\omega) = S(Q,\hbar\omega, T_{base}) - S(Q,\hbar\omega, 300\,\mathrm{K})$ shows thermal population of acoustic phonons as negative intensity, with localized features attributed to sample-environment multiple scattering and additional magnetic modes near 5\,meV and 7\,meV. For $E_i = 100$\,meV [Fig.~\ref{fig:INS-2d}(d)], horizontal bands appear due to temperature-dependent local vibrations, and phonon population effects are seen as a broad negative background, with a low-$Q$ feature near 20\,meV.
Because these excitations arise from finite-size magnetic clusters embedded in a nonmagnetic matrix, they are spatially localized and thus exhibit well-defined, dispersionless energies. The $Q$-dependent modulations reflect cluster geometry via the magnetic structure factor, offering a direct probe of intra-cluster connectivity.

\section{\label{sec:darriet_appendix}Linear Spin-Wave Theory Modeling of Previously Reported Ba$_3$CaRu$_2$O$_9$ Neutron Spectroscopy}

\begin{figure}[h]
\includegraphics[width=\columnwidth]{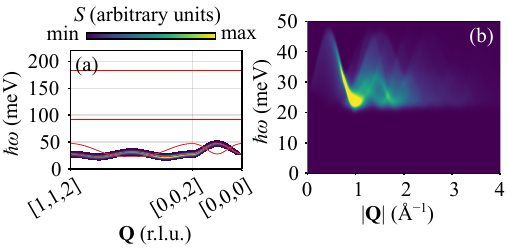}
\caption{(Color online)
\label{fig:BCRO_LSWT_predicted}Simulated inelastic neutron spectra for Ba$_3$CaRu$_2$O$_9$ (BCRO) using best-fit parameters from entangled-units linear spin-wave theory. (a) Simulated single-crystal spectrum $S(\mathbf{Q},\hbar\omega)$ with overlaid magnon modes. (b) Powder-averaged spectrum $S(|\mathbf{Q}|,\hbar\omega)$ showing the gapped triplon band.}
\end{figure}

Additional simulations were performed for BCRO beyond the experimentally available data, Figure~\ref{fig:BCRO_LSWT_predicted}. A single-crystal simulation shown in Fig.~\ref{fig:BCRO_LSWT_predicted}(a) reveals dispersive excitations, consistent with weakly interacting dimers. Figure~\ref{fig:BCRO_LSWT_predicted}(b) shows the powder-averaged dynamic structure factor $S(|\mathbf{Q}|,\hbar\omega)$ computed using the best-fit parameters from entangled-units linear spin-wave theory. A gapped triplon band is clearly visible, exhibiting finite bandwidth.

In the absence of a full resolution function, the instrumental resolution was modeled as a Gaussian with full width at half maximum (FWHM) of 2.7\,meV, matching the reported elastic line width. Simulated spectra were computed by binning $\pm 0.01$\,\AA$^{-1}$ around the reported $Q$ values; alternate bin widths were tested and found to have negligible effect on the fit quality. A single scale factor was used across all $Q$ values, with no fitted background.

The best-fit parameters (Table~\ref{tab:BCRO_fit}) were obtained via a two-stage optimization. First, the exchange constants were initialized near values predicted by DFT, and the Nelder–Mead simplex algorithm was used to find a local minimum. This solution was then refined using the L-BFGS algorithm to ensure convergence. Parameter uncertainties were estimated from the inverse of the Hessian of the loss function (sum of squared residuals), calculated using finite differences with the \texttt{FiniteDiff.jl} package. The covariance matrix was scaled by the experimental variance to obtain standard deviations, $\sigma^2=$

\begin{equation}
\label{eq:cov-resized}
\begin{split}
\resizebox{\columnwidth}{!}{$\displaystyle
\bordermatrix{
   & J_1           & J_2           & J_3           & \text{scale} \cr
J_1     & 1.07\times10^{-1} & 1.64\times10^{-2} & 8.68\times10^{-3} & 5.11\times10^{-4} \cr
J_2     & 1.64\times10^{-2} & 5.67\times10^{-3} & 4.57\times10^{-5} & 1.40\times10^{-4} \cr
J_3     & 8.68\times10^{-3} & 4.57\times10^{-5} & 1.96\times10^{-3} & 4.28\times10^{-5} \cr
\text{scale} & 5.11\times10^{-4} & 1.40\times10^{-4} & 4.28\times10^{-5} & 6.53\times10^{-5}
}
$}
\end{split}
\end{equation}

\section{\label{sec:terasaki_appendix}Dimer Mean-Field Theory Modeling of Previously Reported Ba$_3$Zn$_{1-x}$Ca$_x$Ru$_2$O$_9$ Magnetization Data}

\begin{figure}[hbt]
\includegraphics[width=\columnwidth]{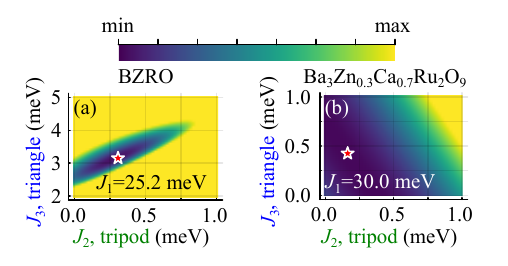}
\caption{(Color online)
\label{fig:loss_landscape}Loss function landscapes for fits to magnetic susceptibility using finite-temperature dimer mean-field theory with $S = \tfrac{3}{2}$. Normalized fitting error is shown as a function of interdimer exchange parameters $J_2$ and $J_3$ for (a) BZRO (color limits: 0–350), and (b) Ba$_3$Zn$_{0.7}$Ca$_{0.3}$Ru$_2$O$_9$ (color limits: 150–350). The intradimer coupling $J_1$ was fixed at the value indicated in each panel. Red stars denote the best-fit parameters reported in Table~\ref{tab:Terasaki_fit_params}.}
\end{figure}

To illustrate the quality of the solutions obtained from fitting the magnetic susceptibility, loss function maps are presented in Fig.~\ref{fig:loss_landscape}. The intradimer exchange $J_1$ was fixed at its best-fit value, while the interdimer couplings $J_2$ and $J_3$ were varied systematically. 

For BZRO, the loss function exhibits a narrow, well-defined valley in the ($J_2$, $J_3$) parameter space with a positive covariance, reflecting the strong influence of interdimer interactions on the magnetic response due to an admixed ground state with finite moment. In contrast, the broader and flatter minimum observed for Ba$_3$Zn$_{0.7}$Ca$_{0.3}$Ru$_2$O$_9$ reflects the relative insensitivity of the susceptibility to the details of the interdimer interactions in a singlet ground state system, as evidenced by the negative covariance.

\clearpage

\nocite{*}

\bibliographystyle{apsrev4-2}
\bibliography{refs}% Produces the bibliography via BibTeX.

\begin{thebibliography}{35}%
\makeatletter
\providecommand \@ifxundefined [1]{%
 \@ifx{#1\undefined}
}%
\providecommand \@ifnum [1]{%
 \ifnum #1\expandafter \@firstoftwo
 \else \expandafter \@secondoftwo
 \fi
}%
\providecommand \@ifx [1]{%
 \ifx #1\expandafter \@firstoftwo
 \else \expandafter \@secondoftwo
 \fi
}%
\providecommand \natexlab [1]{#1}%
\providecommand \enquote  [1]{``#1''}%
\providecommand \bibnamefont  [1]{#1}%
\providecommand \bibfnamefont [1]{#1}%
\providecommand \citenamefont [1]{#1}%
\providecommand \href@noop [0]{\@secondoftwo}%
\providecommand \href [0]{\begingroup \@sanitize@url \@href}%
\providecommand \@href[1]{\@@startlink{#1}\@@href}%
\providecommand \@@href[1]{\endgroup#1\@@endlink}%
\providecommand \@sanitize@url [0]{\catcode `\\12\catcode `\$12\catcode
  `\&12\catcode `\#12\catcode `\^12\catcode `\_12\catcode `\%12\relax}%
\providecommand \@@startlink[1]{}%
\providecommand \@@endlink[0]{}%
\providecommand \url  [0]{\begingroup\@sanitize@url \@url }%
\providecommand \@url [1]{\endgroup\@href {#1}{\urlprefix }}%
\providecommand \urlprefix  [0]{URL }%
\providecommand \Eprint [0]{\href }%
\providecommand \doibase [0]{https://doi.org/}%
\providecommand \selectlanguage [0]{\@gobble}%
\providecommand \bibinfo  [0]{\@secondoftwo}%
\providecommand \bibfield  [0]{\@secondoftwo}%
\providecommand \translation [1]{[#1]}%
\providecommand \BibitemOpen [0]{}%
\providecommand \bibitemStop [0]{}%
\providecommand \bibitemNoStop [0]{.\EOS\space}%
\providecommand \EOS [0]{\spacefactor3000\relax}%
\providecommand \BibitemShut  [1]{\csname bibitem#1\endcsname}%
\let\auto@bib@innerbib\@empty
%</preamble>
\bibitem [{\citenamefont {De}\ \emph {et~al.}(2024)\citenamefont {De},
  \citenamefont {Kurian}, \citenamefont {Solaiappan}, \citenamefont {Kumar},
  \citenamefont {Das}, \citenamefont {Samanta}, \citenamefont {Arafath},\ and\
  \citenamefont {Nair}}]{De2024}%
  \BibitemOpen
  \bibfield  {author} {\bibinfo {author} {\bibfnamefont {S.}~\bibnamefont
  {De}}, \bibinfo {author} {\bibfnamefont {M.~M.}\ \bibnamefont {Kurian}},
  \bibinfo {author} {\bibfnamefont {L.}~\bibnamefont {Solaiappan}}, \bibinfo
  {author} {\bibfnamefont {A.}~\bibnamefont {Kumar}}, \bibinfo {author}
  {\bibfnamefont {S.}~\bibnamefont {Das}}, \bibinfo {author} {\bibfnamefont
  {S.}~\bibnamefont {Samanta}}, \bibinfo {author} {\bibfnamefont {Y.~K.}\
  \bibnamefont {Arafath}},\ and\ \bibinfo {author} {\bibfnamefont {S.~P.~N.}\
  \bibnamefont {Nair}},\ }\href {https://doi.org/10.1021/acs.chemmater.4c00603}
  {\bibfield  {journal} {\bibinfo  {journal} {Chem. Mater.}\ }\textbf {\bibinfo
  {volume} {36}},\ \bibinfo {pages} {2179} (\bibinfo {year}
  {2024})}\BibitemShut {NoStop}%
\bibitem [{\citenamefont {Burbank}\ and\ \citenamefont
  {Evans}(1948)}]{Burbank1948}%
  \BibitemOpen
  \bibfield  {author} {\bibinfo {author} {\bibfnamefont {R.~D.}\ \bibnamefont
  {Burbank}}\ and\ \bibinfo {author} {\bibfnamefont {H.~T.}\ \bibnamefont
  {Evans}},\ }\href {https://doi.org/10.1107/S0365110X48000867} {\bibfield
  {journal} {\bibinfo  {journal} {Acta Crystallogr.}\ }\textbf {\bibinfo
  {volume} {1}},\ \bibinfo {pages} {330} (\bibinfo {year} {1948})}\BibitemShut
  {NoStop}%
\bibitem [{\citenamefont {Shannon}(1976)}]{Shannon1976}%
  \BibitemOpen
  \bibfield  {author} {\bibinfo {author} {\bibfnamefont {R.~D.}\ \bibnamefont
  {Shannon}},\ }\href {https://doi.org/10.1107/S0567739476001551} {\bibfield
  {journal} {\bibinfo  {journal} {Acta Crystallogr. A}\ }\textbf {\bibinfo
  {volume} {32}},\ \bibinfo {pages} {751} (\bibinfo {year} {1976})}\BibitemShut
  {NoStop}%
\bibitem [{\citenamefont {Terasaki}\ \emph {et~al.}(2017)\citenamefont
  {Terasaki}, \citenamefont {Igarashi}, \citenamefont {Nagai}, \citenamefont
  {Tanabe}, \citenamefont {Taniguchi}, \citenamefont {Matsushita},
  \citenamefont {Wada}, \citenamefont {Takata}, \citenamefont {Kida},
  \citenamefont {Hagiwara}, \citenamefont {Kobayashi}, \citenamefont
  {Sagayama}, \citenamefont {Kumai}, \citenamefont {Nakao},\ and\ \citenamefont
  {Murakami}}]{Terasaki2017}%
  \BibitemOpen
  \bibfield  {author} {\bibinfo {author} {\bibfnamefont {I.}~\bibnamefont
  {Terasaki}}, \bibinfo {author} {\bibfnamefont {T.}~\bibnamefont {Igarashi}},
  \bibinfo {author} {\bibfnamefont {T.}~\bibnamefont {Nagai}}, \bibinfo
  {author} {\bibfnamefont {K.}~\bibnamefont {Tanabe}}, \bibinfo {author}
  {\bibfnamefont {H.}~\bibnamefont {Taniguchi}}, \bibinfo {author}
  {\bibfnamefont {T.}~\bibnamefont {Matsushita}}, \bibinfo {author}
  {\bibfnamefont {N.}~\bibnamefont {Wada}}, \bibinfo {author} {\bibfnamefont
  {A.}~\bibnamefont {Takata}}, \bibinfo {author} {\bibfnamefont
  {T.}~\bibnamefont {Kida}}, \bibinfo {author} {\bibfnamefont {M.}~\bibnamefont
  {Hagiwara}}, \bibinfo {author} {\bibfnamefont {K.}~\bibnamefont {Kobayashi}},
  \bibinfo {author} {\bibfnamefont {H.}~\bibnamefont {Sagayama}}, \bibinfo
  {author} {\bibfnamefont {R.}~\bibnamefont {Kumai}}, \bibinfo {author}
  {\bibfnamefont {H.}~\bibnamefont {Nakao}},\ and\ \bibinfo {author}
  {\bibfnamefont {Y.}~\bibnamefont {Murakami}},\ }\href
  {https://doi.org/10.7566/JPSJ.86.033702} {\bibfield  {journal} {\bibinfo
  {journal} {J. Phys. Soc. Jpn.}\ }\textbf {\bibinfo {volume} {86}},\ \bibinfo
  {pages} {033702} (\bibinfo {year} {2017})}\BibitemShut {NoStop}%
\bibitem [{\citenamefont {Darriet}\ \emph {et~al.}(1976)\citenamefont
  {Darriet}, \citenamefont {Drillon}, \citenamefont {Villeneuve},\ and\
  \citenamefont {Hagenmuller}}]{Darriet1976}%
  \BibitemOpen
  \bibfield  {author} {\bibinfo {author} {\bibfnamefont {J.}~\bibnamefont
  {Darriet}}, \bibinfo {author} {\bibfnamefont {M.}~\bibnamefont {Drillon}},
  \bibinfo {author} {\bibfnamefont {G.}~\bibnamefont {Villeneuve}},\ and\
  \bibinfo {author} {\bibfnamefont {P.}~\bibnamefont {Hagenmuller}},\ }\href
  {https://doi.org/10.1016/0022-4596(76)90170-5} {\bibfield  {journal}
  {\bibinfo  {journal} {J. Solid State Chem.}\ }\textbf {\bibinfo {volume}
  {19}},\ \bibinfo {pages} {213} (\bibinfo {year} {1976})}\BibitemShut
  {NoStop}%
\bibitem [{\citenamefont {Drillon}(1977)}]{Drillon1977}%
  \BibitemOpen
  \bibfield  {author} {\bibinfo {author} {\bibfnamefont {M.}~\bibnamefont
  {Drillon}},\ }\href {https://doi.org/10.1016/0038-1098(77)91339-2} {\bibfield
   {journal} {\bibinfo  {journal} {Solid State Commun.}\ }\textbf {\bibinfo
  {volume} {21}},\ \bibinfo {pages} {425} (\bibinfo {year} {1977})}\BibitemShut
  {NoStop}%
\bibitem [{\citenamefont {Darriet}\ \emph {et~al.}(1983)\citenamefont
  {Darriet}, \citenamefont {Soubeyroux},\ and\ \citenamefont
  {Murani}}]{Darriet1983}%
  \BibitemOpen
  \bibfield  {author} {\bibinfo {author} {\bibfnamefont {J.}~\bibnamefont
  {Darriet}}, \bibinfo {author} {\bibfnamefont {L.}~\bibnamefont
  {Soubeyroux}},\ and\ \bibinfo {author} {\bibfnamefont {A.~P.}\ \bibnamefont
  {Murani}},\ }\href {https://doi.org/10.1016/0022-3697(83)90062-8} {\bibfield
  {journal} {\bibinfo  {journal} {J. Phys. Chem. Solids}\ }\textbf {\bibinfo
  {volume} {44}},\ \bibinfo {pages} {269} (\bibinfo {year} {1983})}\BibitemShut
  {NoStop}%
\bibitem [{\citenamefont {Lightfoot}\ and\ \citenamefont
  {Battle}(1990)}]{Lightfoot1990}%
  \BibitemOpen
  \bibfield  {author} {\bibinfo {author} {\bibfnamefont {P.}~\bibnamefont
  {Lightfoot}}\ and\ \bibinfo {author} {\bibfnamefont {P.~D.}\ \bibnamefont
  {Battle}},\ }\href {https://doi.org/10.1016/0022-4596(90)90309-L} {\bibfield
  {journal} {\bibinfo  {journal} {J. Solid State Chem.}\ }\textbf {\bibinfo
  {volume} {89}},\ \bibinfo {pages} {174} (\bibinfo {year} {1990})}\BibitemShut
  {NoStop}%
\bibitem [{\citenamefont {Fernandez}\ \emph {et~al.}(1980)\citenamefont
  {Fernandez}, \citenamefont {Greatrex},\ and\ \citenamefont
  {Greenwood}}]{Fernandez1980}%
  \BibitemOpen
  \bibfield  {author} {\bibinfo {author} {\bibfnamefont {I.}~\bibnamefont
  {Fernandez}}, \bibinfo {author} {\bibfnamefont {R.}~\bibnamefont
  {Greatrex}},\ and\ \bibinfo {author} {\bibfnamefont {N.~N.}\ \bibnamefont
  {Greenwood}},\ }\href {https://doi.org/10.1016/0022-4596(80)90410-7}
  {\bibfield  {journal} {\bibinfo  {journal} {J. Solid State Chem.}\ }\textbf
  {\bibinfo {volume} {34}},\ \bibinfo {pages} {121} (\bibinfo {year}
  {1980})}\BibitemShut {NoStop}%
\bibitem [{\citenamefont {Beran}\ \emph {et~al.}(2015)\citenamefont {Beran},
  \citenamefont {Ivanov}, \citenamefont {Nordblad}, \citenamefont {Middey},
  \citenamefont {Nag}, \citenamefont {Sarma}, \citenamefont {Ray},\ and\
  \citenamefont {Mathieu}}]{Beran2015}%
  \BibitemOpen
  \bibfield  {author} {\bibinfo {author} {\bibfnamefont {P.}~\bibnamefont
  {Beran}}, \bibinfo {author} {\bibfnamefont {S.~A.}\ \bibnamefont {Ivanov}},
  \bibinfo {author} {\bibfnamefont {P.}~\bibnamefont {Nordblad}}, \bibinfo
  {author} {\bibfnamefont {S.}~\bibnamefont {Middey}}, \bibinfo {author}
  {\bibfnamefont {A.}~\bibnamefont {Nag}}, \bibinfo {author} {\bibfnamefont
  {D.~D.}\ \bibnamefont {Sarma}}, \bibinfo {author} {\bibfnamefont
  {S.}~\bibnamefont {Ray}},\ and\ \bibinfo {author} {\bibfnamefont
  {R.}~\bibnamefont {Mathieu}},\ }\href
  {https://doi.org/10.1016/j.solidstatesciences.2015.10.011} {\bibfield
  {journal} {\bibinfo  {journal} {Solid State Sci.}\ }\textbf {\bibinfo
  {volume} {50}},\ \bibinfo {pages} {58} (\bibinfo {year} {2015})}\BibitemShut
  {NoStop}%
\bibitem [{\citenamefont {Senn}\ \emph {et~al.}(2013)\citenamefont {Senn},
  \citenamefont {Arevalo-Lopez}, \citenamefont {Saito}, \citenamefont
  {Shimakawa},\ and\ \citenamefont {Attfield}}]{Senn2013}%
  \BibitemOpen
  \bibfield  {author} {\bibinfo {author} {\bibfnamefont {M.~S.}\ \bibnamefont
  {Senn}}, \bibinfo {author} {\bibfnamefont {A.~M.}\ \bibnamefont
  {Arevalo-Lopez}}, \bibinfo {author} {\bibfnamefont {T.}~\bibnamefont
  {Saito}}, \bibinfo {author} {\bibfnamefont {Y.}~\bibnamefont {Shimakawa}},\
  and\ \bibinfo {author} {\bibfnamefont {J.~P.}\ \bibnamefont {Attfield}},\
  }\href {https://doi.org/10.1088/0953-8984/25/49/496008} {\bibfield  {journal}
  {\bibinfo  {journal} {J. Phys.: Condens. Matter}\ }\textbf {\bibinfo {volume}
  {25}},\ \bibinfo {pages} {496008} (\bibinfo {year} {2013})}\BibitemShut
  {NoStop}%
\bibitem [{\citenamefont {Yamamoto}\ \emph {et~al.}(2018)\citenamefont
  {Yamamoto}, \citenamefont {Taniguchi},\ and\ \citenamefont
  {Terasaki}}]{Yamamoto2018}%
  \BibitemOpen
  \bibfield  {author} {\bibinfo {author} {\bibfnamefont {T.~D.}\ \bibnamefont
  {Yamamoto}}, \bibinfo {author} {\bibfnamefont {H.}~\bibnamefont
  {Taniguchi}},\ and\ \bibinfo {author} {\bibfnamefont {I.}~\bibnamefont
  {Terasaki}},\ }\href {https://doi.org/10.1088/1361-648X/aad5ac} {\bibfield
  {journal} {\bibinfo  {journal} {J. Phys.: Condens. Matter}\ }\textbf
  {\bibinfo {volume} {30}},\ \bibinfo {pages} {355802} (\bibinfo {year}
  {2018})}\BibitemShut {NoStop}%
\bibitem [{\citenamefont {Nishihara}\ \emph {et~al.}(2021)\citenamefont
  {Nishihara}, \citenamefont {Kimura}, \citenamefont {Nakano}, \citenamefont
  {Taniguchi},\ and\ \citenamefont {Terasaki}}]{Nishihara2021}%
  \BibitemOpen
  \bibfield  {author} {\bibinfo {author} {\bibfnamefont {D.}~\bibnamefont
  {Nishihara}}, \bibinfo {author} {\bibfnamefont {A.}~\bibnamefont {Kimura}},
  \bibinfo {author} {\bibfnamefont {A.}~\bibnamefont {Nakano}}, \bibinfo
  {author} {\bibfnamefont {H.}~\bibnamefont {Taniguchi}},\ and\ \bibinfo
  {author} {\bibfnamefont {I.}~\bibnamefont {Terasaki}},\ }\href
  {https://doi.org/10.7566/JPSJ.90.124707} {\bibfield  {journal} {\bibinfo
  {journal} {J. Phys. Soc. Jpn.}\ }\textbf {\bibinfo {volume} {90}},\ \bibinfo
  {pages} {124707} (\bibinfo {year} {2021})}\BibitemShut {NoStop}%
\bibitem [{\citenamefont {Ochiai}\ \emph {et~al.}(2024)\citenamefont {Ochiai},
  \citenamefont {Terasaki},\ and\ \citenamefont {Yasui}}]{Ochiai2024}%
  \BibitemOpen
  \bibfield  {author} {\bibinfo {author} {\bibfnamefont {Y.}~\bibnamefont
  {Ochiai}}, \bibinfo {author} {\bibfnamefont {I.}~\bibnamefont {Terasaki}},\
  and\ \bibinfo {author} {\bibfnamefont {Y.}~\bibnamefont {Yasui}},\ }\href
  {https://doi.org/10.1063/9.0000667} {\bibfield  {journal} {\bibinfo
  {journal} {AIP Adv.}\ }\textbf {\bibinfo {volume} {14}},\ \bibinfo {pages}
  {025126} (\bibinfo {year} {2024})}\BibitemShut {NoStop}%
\bibitem [{\citenamefont {Hayashida}\ \emph {et~al.}(2025)\citenamefont
  {Hayashida}, \citenamefont {Gretarsson}, \citenamefont {Puphal},
  \citenamefont {Isobe}, \citenamefont {Goering}, \citenamefont {Matsumoto},
  \citenamefont {Nuss}, \citenamefont {Takagi}, \citenamefont {Hepting},\ and\
  \citenamefont {Keimer}}]{Hayashida2025}%
  \BibitemOpen
  \bibfield  {author} {\bibinfo {author} {\bibfnamefont {S.}~\bibnamefont
  {Hayashida}}, \bibinfo {author} {\bibfnamefont {H.}~\bibnamefont
  {Gretarsson}}, \bibinfo {author} {\bibfnamefont {P.}~\bibnamefont {Puphal}},
  \bibinfo {author} {\bibfnamefont {M.}~\bibnamefont {Isobe}}, \bibinfo
  {author} {\bibfnamefont {E.}~\bibnamefont {Goering}}, \bibinfo {author}
  {\bibfnamefont {Y.}~\bibnamefont {Matsumoto}}, \bibinfo {author}
  {\bibfnamefont {J.}~\bibnamefont {Nuss}}, \bibinfo {author} {\bibfnamefont
  {H.}~\bibnamefont {Takagi}}, \bibinfo {author} {\bibfnamefont
  {M.}~\bibnamefont {Hepting}},\ and\ \bibinfo {author} {\bibfnamefont
  {B.}~\bibnamefont {Keimer}},\ }\href
  {https://doi.org/10.1103/PhysRevB.111.104418} {\bibfield  {journal} {\bibinfo
   {journal} {Phys. Rev. B}\ }\textbf {\bibinfo {volume} {111}},\ \bibinfo
  {pages} {104418} (\bibinfo {year} {2025})}\BibitemShut {NoStop}%
\bibitem [{\citenamefont {White}(2007)}]{White2007}%
  \BibitemOpen
  \bibfield  {author} {\bibinfo {author} {\bibfnamefont {R.~M.}\ \bibnamefont
  {White}},\ }\href@noop {} {\emph {\bibinfo {title} {{Quantum Theory of
  Magnetism}}}},\ \bibinfo {edition} {3rd}\ ed.\ (\bibinfo  {publisher}
  {Springer},\ \bibinfo {address} {Berlin},\ \bibinfo {year}
  {2007})\BibitemShut {NoStop}%
\bibitem [{\citenamefont {Dahlbom}\ \emph {et~al.}(2024)\citenamefont
  {Dahlbom}, \citenamefont {Thomas}, \citenamefont {Johnston}, \citenamefont
  {Barros},\ and\ \citenamefont {Batista}}]{Dahlbom2024}%
  \BibitemOpen
  \bibfield  {author} {\bibinfo {author} {\bibfnamefont {D.~A.}\ \bibnamefont
  {Dahlbom}}, \bibinfo {author} {\bibfnamefont {J.}~\bibnamefont {Thomas}},
  \bibinfo {author} {\bibfnamefont {S.}~\bibnamefont {Johnston}}, \bibinfo
  {author} {\bibfnamefont {K.}~\bibnamefont {Barros}},\ and\ \bibinfo {author}
  {\bibfnamefont {C.~D.}\ \bibnamefont {Batista}},\ }\href
  {https://doi.org/10.1103/PhysRevB.110.104403} {\bibfield  {journal} {\bibinfo
   {journal} {Phys. Rev. B}\ }\textbf {\bibinfo {volume} {110}},\ \bibinfo
  {pages} {104403} (\bibinfo {year} {2024})}\BibitemShut {NoStop}%
\bibitem [{\citenamefont {Dahlbom}\ \emph {et~al.}(2025)\citenamefont
  {Dahlbom}, \citenamefont {Zhang}, \citenamefont {Miles}, \citenamefont
  {Quinn}, \citenamefont {Niraula}, \citenamefont {Thipe}, \citenamefont
  {Wilson}, \citenamefont {Matin}, \citenamefont {Mankad}, \citenamefont
  {Hahn}, \citenamefont {Pajerowski}, \citenamefont {Johnston}, \citenamefont
  {Wang}, \citenamefont {Lane}, \citenamefont {Li}, \citenamefont {Bai},
  \citenamefont {Mourigal}, \citenamefont {Batista},\ and\ \citenamefont
  {Barros}}]{Dahlbom2025}%
  \BibitemOpen
  \bibfield  {author} {\bibinfo {author} {\bibfnamefont {D.}~\bibnamefont
  {Dahlbom}}, \bibinfo {author} {\bibfnamefont {H.}~\bibnamefont {Zhang}},
  \bibinfo {author} {\bibfnamefont {C.}~\bibnamefont {Miles}}, \bibinfo
  {author} {\bibfnamefont {S.}~\bibnamefont {Quinn}}, \bibinfo {author}
  {\bibfnamefont {A.}~\bibnamefont {Niraula}}, \bibinfo {author} {\bibfnamefont
  {B.}~\bibnamefont {Thipe}}, \bibinfo {author} {\bibfnamefont
  {M.}~\bibnamefont {Wilson}}, \bibinfo {author} {\bibfnamefont
  {S.}~\bibnamefont {Matin}}, \bibinfo {author} {\bibfnamefont
  {H.}~\bibnamefont {Mankad}}, \bibinfo {author} {\bibfnamefont
  {S.}~\bibnamefont {Hahn}}, \bibinfo {author} {\bibfnamefont {D.}~\bibnamefont
  {Pajerowski}}, \bibinfo {author} {\bibfnamefont {S.}~\bibnamefont
  {Johnston}}, \bibinfo {author} {\bibfnamefont {Z.}~\bibnamefont {Wang}},
  \bibinfo {author} {\bibfnamefont {H.}~\bibnamefont {Lane}}, \bibinfo {author}
  {\bibfnamefont {Y.~W.}\ \bibnamefont {Li}}, \bibinfo {author} {\bibfnamefont
  {X.}~\bibnamefont {Bai}}, \bibinfo {author} {\bibfnamefont {M.}~\bibnamefont
  {Mourigal}}, \bibinfo {author} {\bibfnamefont {C.~D.}\ \bibnamefont
  {Batista}},\ and\ \bibinfo {author} {\bibfnamefont {K.}~\bibnamefont
  {Barros}},\ }\href@noop {} {\bibinfo {title} {{Sunny.jl: A Julia package for
  spin dynamics}}} (\bibinfo {year} {2025}),\ \Eprint
  {https://arxiv.org/abs/2501.13095} {arXiv:2501.13095 [quant-ph]} \BibitemShut
  {NoStop}%
\bibitem [{\citenamefont {Toth}\ and\ \citenamefont {Lake}(2015)}]{Toth2015}%
  \BibitemOpen
  \bibfield  {author} {\bibinfo {author} {\bibfnamefont {S.}~\bibnamefont
  {Toth}}\ and\ \bibinfo {author} {\bibfnamefont {B.}~\bibnamefont {Lake}},\
  }\href {https://doi.org/10.1088/0953-8984/27/16/166002} {\bibfield  {journal}
  {\bibinfo  {journal} {J. Phys.: Condens. Matter}\ }\textbf {\bibinfo {volume}
  {27}},\ \bibinfo {pages} {166002} (\bibinfo {year} {2015})}\BibitemShut
  {NoStop}%
\bibitem [{\citenamefont {Zhang}\ \emph {et~al.}(2025)\citenamefont {Zhang},
  \citenamefont {Kato}, \citenamefont {Ghioldi}, \citenamefont {Manuel},
  \citenamefont {Trumper},\ and\ \citenamefont {Batista}}]{Zhang2025}%
  \BibitemOpen
  \bibfield  {author} {\bibinfo {author} {\bibfnamefont {S.-S.}\ \bibnamefont
  {Zhang}}, \bibinfo {author} {\bibfnamefont {Y.}~\bibnamefont {Kato}},
  \bibinfo {author} {\bibfnamefont {E.~A.}\ \bibnamefont {Ghioldi}}, \bibinfo
  {author} {\bibfnamefont {L.~O.}\ \bibnamefont {Manuel}}, \bibinfo {author}
  {\bibfnamefont {A.~E.}\ \bibnamefont {Trumper}},\ and\ \bibinfo {author}
  {\bibfnamefont {C.~D.}\ \bibnamefont {Batista}},\ }\href
  {https://doi.org/10.1103/qfmw-whbp} {\bibfield  {journal} {\bibinfo
  {journal} {Phys. Rev. B}\ }\textbf {\bibinfo {volume} {112}},\ \bibinfo
  {pages} {024410} (\bibinfo {year} {2025})}\BibitemShut {NoStop}%
\bibitem [{\citenamefont {Streltsov}(2013)}]{Streltsov2013}%
  \BibitemOpen
  \bibfield  {author} {\bibinfo {author} {\bibfnamefont {S.~V.}\ \bibnamefont
  {Streltsov}},\ }\href {https://doi.org/10.1103/PhysRevB.88.024429} {\bibfield
   {journal} {\bibinfo  {journal} {Phys. Rev. B}\ }\textbf {\bibinfo {volume}
  {88}},\ \bibinfo {pages} {024429} (\bibinfo {year} {2013})},\ \Eprint
  {https://arxiv.org/abs/1306.3333} {arXiv:1306.3333} \BibitemShut {NoStop}%
\bibitem [{\citenamefont {Furrer}\ \emph {et~al.}(2011)\citenamefont {Furrer},
  \citenamefont {Pomjakushina}, \citenamefont {Pomjakushin}, \citenamefont
  {Embs},\ and\ \citenamefont {Str{\"a}ssle}}]{Furrer2011}%
  \BibitemOpen
  \bibfield  {author} {\bibinfo {author} {\bibfnamefont {A.}~\bibnamefont
  {Furrer}}, \bibinfo {author} {\bibfnamefont {E.}~\bibnamefont
  {Pomjakushina}}, \bibinfo {author} {\bibfnamefont {V.}~\bibnamefont
  {Pomjakushin}}, \bibinfo {author} {\bibfnamefont {J.~P.}\ \bibnamefont
  {Embs}},\ and\ \bibinfo {author} {\bibfnamefont {T.}~\bibnamefont
  {Str{\"a}ssle}},\ }\href {https://doi.org/10.1103/PhysRevB.83.174442}
  {\bibfield  {journal} {\bibinfo  {journal} {Phys. Rev. B}\ }\textbf {\bibinfo
  {volume} {83}},\ \bibinfo {pages} {174442} (\bibinfo {year}
  {2011})}\BibitemShut {NoStop}%
\bibitem [{\citenamefont {Tanaka}\ and\ \citenamefont
  {Hotta}(2020)}]{Tanaka2020}%
  \BibitemOpen
  \bibfield  {author} {\bibinfo {author} {\bibfnamefont {K.}~\bibnamefont
  {Tanaka}}\ and\ \bibinfo {author} {\bibfnamefont {C.}~\bibnamefont {Hotta}},\
  }\href {https://doi.org/10.1103/PhysRevB.101.094422} {\bibfield  {journal}
  {\bibinfo  {journal} {Phys. Rev. B}\ }\textbf {\bibinfo {volume} {101}},\
  \bibinfo {pages} {094422} (\bibinfo {year} {2020})}\BibitemShut {NoStop}%
\bibitem [{\citenamefont {Barker}\ and\ \citenamefont
  {Bauer}(2019)}]{barker2019}%
  \BibitemOpen
  \bibfield  {author} {\bibinfo {author} {\bibfnamefont {J.}~\bibnamefont
  {Barker}}\ and\ \bibinfo {author} {\bibfnamefont {G.~E.}\ \bibnamefont
  {Bauer}},\ }\href@noop {} {\bibfield  {journal} {\bibinfo  {journal} {Phys.
  Rev. B}\ }\textbf {\bibinfo {volume} {100}},\ \bibinfo {pages} {140401}
  (\bibinfo {year} {2019})}\BibitemShut {NoStop}%
\bibitem [{\citenamefont {Williams}\ \emph {et~al.}(2025)\citenamefont
  {Williams}, \citenamefont {Dahlbom}, \citenamefont {Zhang}, \citenamefont
  {Agarwal}, \citenamefont {Barros},\ and\ \citenamefont
  {Batista}}]{Williams2025}%
  \BibitemOpen
  \bibfield  {author} {\bibinfo {author} {\bibfnamefont {F.}~\bibnamefont
  {Williams}}, \bibinfo {author} {\bibfnamefont {D.}~\bibnamefont {Dahlbom}},
  \bibinfo {author} {\bibfnamefont {H.}~\bibnamefont {Zhang}}, \bibinfo
  {author} {\bibfnamefont {S.}~\bibnamefont {Agarwal}}, \bibinfo {author}
  {\bibfnamefont {K.}~\bibnamefont {Barros}},\ and\ \bibinfo {author}
  {\bibfnamefont {C.~D.}\ \bibnamefont {Batista}},\ }\href@noop {} {\bibinfo
  {title} {{Skyrmions of frustrated quantum dimer systems}}} (\bibinfo {year}
  {2025}),\ \Eprint {https://arxiv.org/abs/2506.22320} {arXiv:2506.22320
  [cond-mat.str-el]} \BibitemShut {NoStop}%
\bibitem [{\citenamefont {Luttinger}\ and\ \citenamefont
  {Tisza}(1946)}]{Luttinger1946}%
  \BibitemOpen
  \bibfield  {author} {\bibinfo {author} {\bibfnamefont {J.~M.}\ \bibnamefont
  {Luttinger}}\ and\ \bibinfo {author} {\bibfnamefont {L.}~\bibnamefont
  {Tisza}},\ }\href {https://doi.org/10.1103/PhysRev.70.954} {\bibfield
  {journal} {\bibinfo  {journal} {Phys. Rev.}\ }\textbf {\bibinfo {volume}
  {70}},\ \bibinfo {pages} {954} (\bibinfo {year} {1946})}\BibitemShut
  {NoStop}%
\bibitem [{\citenamefont {Litvin}(1974)}]{Litvin1974}%
  \BibitemOpen
  \bibfield  {author} {\bibinfo {author} {\bibfnamefont {D.~B.}\ \bibnamefont
  {Litvin}},\ }\href {https://doi.org/10.1016/0031-8914(74)90243-2} {\bibfield
  {journal} {\bibinfo  {journal} {Physica}\ }\textbf {\bibinfo {volume} {77}},\
  \bibinfo {pages} {205} (\bibinfo {year} {1974})}\BibitemShut {NoStop}%
\bibitem [{\citenamefont {Niemeijer}\ and\ \citenamefont
  {Bl{\"o}te}(1973)}]{Niemeijer1973}%
  \BibitemOpen
  \bibfield  {author} {\bibinfo {author} {\bibfnamefont {T.}~\bibnamefont
  {Niemeijer}}\ and\ \bibinfo {author} {\bibfnamefont {H.~W.~J.}\ \bibnamefont
  {Bl{\"o}te}},\ }\href {https://doi.org/10.1016/0031-8914(73)90055-3}
  {\bibfield  {journal} {\bibinfo  {journal} {Physica}\ }\textbf {\bibinfo
  {volume} {67}},\ \bibinfo {pages} {125} (\bibinfo {year} {1973})}\BibitemShut
  {NoStop}%
\bibitem [{\citenamefont {Kresse}\ and\ \citenamefont
  {Furthm{\"u}ller}(1996{\natexlab{a}})}]{Kresse1996_CMS}%
  \BibitemOpen
  \bibfield  {author} {\bibinfo {author} {\bibfnamefont {G.}~\bibnamefont
  {Kresse}}\ and\ \bibinfo {author} {\bibfnamefont {J.}~\bibnamefont
  {Furthm{\"u}ller}},\ }\href {https://doi.org/10.1016/0927-0256(96)00008-0}
  {\bibfield  {journal} {\bibinfo  {journal} {Comput. Mater. Sci.}\ }\textbf
  {\bibinfo {volume} {6}},\ \bibinfo {pages} {15} (\bibinfo {year}
  {1996}{\natexlab{a}})}\BibitemShut {NoStop}%
\bibitem [{\citenamefont {Kresse}\ and\ \citenamefont
  {Furthm{\"u}ller}(1996{\natexlab{b}})}]{Kresse1996_PRB}%
  \BibitemOpen
  \bibfield  {author} {\bibinfo {author} {\bibfnamefont {G.}~\bibnamefont
  {Kresse}}\ and\ \bibinfo {author} {\bibfnamefont {J.}~\bibnamefont
  {Furthm{\"u}ller}},\ }\href {https://doi.org/10.1103/PhysRevB.54.11169}
  {\bibfield  {journal} {\bibinfo  {journal} {Phys. Rev. B}\ }\textbf {\bibinfo
  {volume} {54}},\ \bibinfo {pages} {11169} (\bibinfo {year}
  {1996}{\natexlab{b}})}\BibitemShut {NoStop}%
\bibitem [{\citenamefont {Bl{\"o}chl}(1994)}]{Bloechl1994}%
  \BibitemOpen
  \bibfield  {author} {\bibinfo {author} {\bibfnamefont {P.~E.}\ \bibnamefont
  {Bl{\"o}chl}},\ }\href {https://doi.org/10.1103/PhysRevB.50.17953} {\bibfield
   {journal} {\bibinfo  {journal} {Phys. Rev. B}\ }\textbf {\bibinfo {volume}
  {50}},\ \bibinfo {pages} {17953} (\bibinfo {year} {1994})}\BibitemShut
  {NoStop}%
\bibitem [{\citenamefont {Kresse}\ and\ \citenamefont
  {Joubert}(1999)}]{Kresse1999}%
  \BibitemOpen
  \bibfield  {author} {\bibinfo {author} {\bibfnamefont {G.}~\bibnamefont
  {Kresse}}\ and\ \bibinfo {author} {\bibfnamefont {D.}~\bibnamefont
  {Joubert}},\ }\href {https://doi.org/10.1103/PhysRevB.59.1758} {\bibfield
  {journal} {\bibinfo  {journal} {Phys. Rev. B}\ }\textbf {\bibinfo {volume}
  {59}},\ \bibinfo {pages} {1758} (\bibinfo {year} {1999})}\BibitemShut
  {NoStop}%
\bibitem [{\citenamefont {Perdew}\ \emph {et~al.}(1996)\citenamefont {Perdew},
  \citenamefont {Burke},\ and\ \citenamefont {Ernzerhof}}]{Perdew1996}%
  \BibitemOpen
  \bibfield  {author} {\bibinfo {author} {\bibfnamefont {J.~P.}\ \bibnamefont
  {Perdew}}, \bibinfo {author} {\bibfnamefont {K.}~\bibnamefont {Burke}},\ and\
  \bibinfo {author} {\bibfnamefont {M.}~\bibnamefont {Ernzerhof}},\ }\href
  {https://doi.org/10.1103/PhysRevLett.77.3865} {\bibfield  {journal} {\bibinfo
   {journal} {Phys. Rev. Lett.}\ }\textbf {\bibinfo {volume} {77}},\ \bibinfo
  {pages} {3865} (\bibinfo {year} {1996})}\BibitemShut {NoStop}%
\bibitem [{\citenamefont {Furrer}\ and\ \citenamefont
  {G{\"u}del}(1979)}]{Furrer1979}%
  \BibitemOpen
  \bibfield  {author} {\bibinfo {author} {\bibfnamefont {A.}~\bibnamefont
  {Furrer}}\ and\ \bibinfo {author} {\bibfnamefont {H.~U.}\ \bibnamefont
  {G{\"u}del}},\ }\href {https://doi.org/10.1016/0304-8853(79)90134-3}
  {\bibfield  {journal} {\bibinfo  {journal} {J. Magn. Magn. Mater.}\ }\textbf
  {\bibinfo {volume} {14}},\ \bibinfo {pages} {256} (\bibinfo {year}
  {1979})}\BibitemShut {NoStop}%
\bibitem [{\citenamefont {Arnold}\ \emph {et~al.}(2014)\citenamefont {Arnold},
  \citenamefont {Bilheux}, \citenamefont {Borreguero}, \citenamefont {Buts},
  \citenamefont {Campbell}, \citenamefont {Chapon}, \citenamefont {Doucet},
  \citenamefont {Draper}, \citenamefont {Leal}, \citenamefont {Gigg},
  \citenamefont {Lynch}, \citenamefont {Markvardsen}, \citenamefont
  {Mikkelson}, \citenamefont {Mikkelson}, \citenamefont {Miller}, \citenamefont
  {Palmen}, \citenamefont {Parker}, \citenamefont {Passos}, \citenamefont
  {Perring}, \citenamefont {Peterson}, \citenamefont {Ren}, \citenamefont
  {Reuter}, \citenamefont {Savici}, \citenamefont {Taylor}, \citenamefont
  {Taylor}, \citenamefont {Tolchenov}, \citenamefont {Zhou},\ and\
  \citenamefont {Zikovsky}}]{Arnold2014}%
  \BibitemOpen
  \bibfield  {author} {\bibinfo {author} {\bibfnamefont {O.}~\bibnamefont
  {Arnold}}, \bibinfo {author} {\bibfnamefont {J.~C.}\ \bibnamefont {Bilheux}},
  \bibinfo {author} {\bibfnamefont {J.~M.}\ \bibnamefont {Borreguero}},
  \bibinfo {author} {\bibfnamefont {A.}~\bibnamefont {Buts}}, \bibinfo {author}
  {\bibfnamefont {S.~I.}\ \bibnamefont {Campbell}}, \bibinfo {author}
  {\bibfnamefont {L.}~\bibnamefont {Chapon}}, \bibinfo {author} {\bibfnamefont
  {M.}~\bibnamefont {Doucet}}, \bibinfo {author} {\bibfnamefont
  {N.}~\bibnamefont {Draper}}, \bibinfo {author} {\bibfnamefont {R.~F.}\
  \bibnamefont {Leal}}, \bibinfo {author} {\bibfnamefont {M.~A.}\ \bibnamefont
  {Gigg}}, \bibinfo {author} {\bibfnamefont {V.~E.}\ \bibnamefont {Lynch}},
  \bibinfo {author} {\bibfnamefont {A.}~\bibnamefont {Markvardsen}}, \bibinfo
  {author} {\bibfnamefont {D.~J.}\ \bibnamefont {Mikkelson}}, \bibinfo {author}
  {\bibfnamefont {R.~L.}\ \bibnamefont {Mikkelson}}, \bibinfo {author}
  {\bibfnamefont {R.}~\bibnamefont {Miller}}, \bibinfo {author} {\bibfnamefont
  {K.}~\bibnamefont {Palmen}}, \bibinfo {author} {\bibfnamefont
  {P.}~\bibnamefont {Parker}}, \bibinfo {author} {\bibfnamefont
  {G.}~\bibnamefont {Passos}}, \bibinfo {author} {\bibfnamefont {T.~G.}\
  \bibnamefont {Perring}}, \bibinfo {author} {\bibfnamefont {P.~F.}\
  \bibnamefont {Peterson}}, \bibinfo {author} {\bibfnamefont {S.}~\bibnamefont
  {Ren}}, \bibinfo {author} {\bibfnamefont {M.~A.}\ \bibnamefont {Reuter}},
  \bibinfo {author} {\bibfnamefont {A.~T.}\ \bibnamefont {Savici}}, \bibinfo
  {author} {\bibfnamefont {J.~W.}\ \bibnamefont {Taylor}}, \bibinfo {author}
  {\bibfnamefont {R.~J.}\ \bibnamefont {Taylor}}, \bibinfo {author}
  {\bibfnamefont {R.}~\bibnamefont {Tolchenov}}, \bibinfo {author}
  {\bibfnamefont {W.}~\bibnamefont {Zhou}},\ and\ \bibinfo {author}
  {\bibfnamefont {J.}~\bibnamefont {Zikovsky}},\ }\href
  {https://doi.org/10.1016/j.nima.2014.07.029} {\bibfield  {journal} {\bibinfo
  {journal} {Nucl. Instrum. Methods Phys. Res., Sect. A}\ }\textbf {\bibinfo
  {volume} {764}},\ \bibinfo {pages} {156} (\bibinfo {year}
  {2014})}\BibitemShut {NoStop}%
\end{thebibliography}%

\end{document}